\documentclass[11pt,a4paper,twoside]{article}
\usepackage{amsmath,amssymb}
\usepackage{mathrsfs}
\usepackage{graphicx}
\usepackage{cancel}
\def\gtap{\mathrel{ \rlap{\raise 0.511ex \hbox{$>$}}{\lower 0.511ex
   \hbox{$\sim$}}}}
\def\ltap{\mathrel{ \rlap{\raise 0.511ex
   \hbox{$<$}}{\lower 0.511ex \hbox{$\sim$}}}}
\newcommand{\betabeta}{\mbox{$(\beta \beta)_{0 \nu}$}}
\def\sss{\scriptscriptstyle}

\def\eg{\hbox{e.g.}{}}

\def\meff{\mbox{$\left|\langle m\rangle\right|$}}

\def\deltaatm{\mbox{$\Delta m^2_{\mathrm{A}}$}}
\def\eV{\mbox{\rm eV}}
\def\hbeta{${}^3{\rm H}\,\beta$-decay}

\newcommand{\BR}{\text{BR}}
\newcommand{\PMNS}{{\text{PMNS}}}
\renewcommand{\thefootnote}{\fnsymbol{footnote}}
\begin{document}
\begin{titlepage}
\hfill
\vbox{
    \halign{#\hfil        \cr
      SISSA 18/2009/EP \cr
      TUM-HEP-717/09 \cr
      IPMU09-0037\cr
    } % end of \halign
  }  % end of \vbox
  \vspace*{10mm}
  \begin{center}
    {\Large{\bf Neutrinoless Double Beta Decay and 
$H^{\pm\pm}\to {l'}^\pm l^\pm$ Decays 
in the Higgs Triplet Model}}

\vspace*{7mm}

{\ S.~T.~Petcov$^{a,b)}$}\footnote[1]{Also at: Institute of Nuclear Research
  and Nuclear Energy, Bulgarian Academy %
  of Sciences, 1784 Sofia, Bulgaria},%
{\ H.~Sugiyama$^{a)}$}\footnote[2]{E-mail: sugiyama@sissa.it},%
{\ Y.~Takanishi$^{c)}$}\footnote[3]{E-mail: yasutaka.takanishi@ph.tum.de}

\vspace*{0.7cm}

${}^{a)}$~{\it SISSA and INFN-Sezione di Trieste, I-34014 Trieste, Italy.}\\
${}^{b)}$~{\it IPMU, University of Tokyo, Tokyo, Japan.}\\
${}^{c)}$~{\it Physik-Department, Technische Universit{\"a}t M{\"u}nchen,
  D-85747 Garching, Germany.}\\

\vspace*{1cm}
\end{center}
\begin{abstract}
%%%%%%%%%%%%%%%%%%%%%%%%%%%%%%%%%%%%%%%%%%%%%%%%%%%%%%%
  The connection between the neutrinoless double beta ($\betabeta$-)
  decay effective Majorana mass, $|M_{ee}|$, and the branching ratios
  of the decays $H^{\pm\pm}\to l^\pm {l'}^\pm$, $l,{l'} = e,\mu$, of
  the doubly charged Higgs boson $H^{\pm\pm}$ is analysed within the
  Higgs Triplet Model of neutrino mass generation.  We work in the
  version of the model with explicit breaking of the total lepton
  charge conservation, in which $H^{\pm\pm}\to l^\pm {l'}^\pm$,
  $l,{l'} = e,\mu,\tau$, are the dominant decay modes of $H^{\pm\pm}$.
  It is assumed also that $H^{\pm\pm}$ are relatively light so that
  they can be produced at LHC and the branching ratios of interest
  measured.  Taking into account the current and prospective
  uncertainties in the values of the neutrino mixing parameters most
  relevant for the problem studied - the atmospheric neutrino mixing
  angle $\theta_{23}$ and the CHOOZ angle $\theta_{13}$, and allowing
  the lightest neutrino mass and the CP violating Dirac and Majorana
  phases to vary in the intervals $[0, 0.3~\eV]$ and $[0, 2\pi]$,
  respectively, we derive the regions of values of $\BR(H^{\pm\pm}\to
  e^\pm e^\pm)$ and $\BR(H^{\pm\pm}\to e^\pm \mu^\pm)$ for which
  $|M_{ee}| \geq 0.05$ eV, or $|M_{ee}| < 0.05$ eV.  This is done for
  neutrino mass spectrum with normal ordering, inverted ordering and
  in the case when the type of the spectrum is not known, and i)
  without using the possible additional data on $\BR(H^{\pm\pm}\to
  \mu^\pm \mu^\pm)$, ii) using prospective data on $\BR(H^{\pm\pm}\to
  \mu^\pm \mu^\pm)$.  In the latter case results for several values of
  $\BR(H^{\pm\pm}\to \mu^\pm \mu^\pm)$ are presented.
\end{abstract}
\vskip 0.5cm
April 2009
\end{titlepage}
%%%%%%%%%%%%%%%%%%%%%%%%%%%%%%%%%%%%%%%%%%%%%%%%%%%%%%%
\newpage
\renewcommand{\thefootnote}{\arabic{footnote}}
\setcounter{footnote}{0}
\setcounter{page}{1}
%%%%%%%%%%%%%%%%%%%%%%%%%%%%%%%%%%%%%%%%%%%%%%%%%%%%%%%

\section{Introduction}

  Determining the origin of neutrino masses and mixing is one of the
major challenges of future research in neutrino physics. It is well
known that the existence of nonzero neutrino masses can be related to
the presence of more complicated Higgs sector in the Standard Theory,
involving additional Higgs fields beyond the single doublet field.
Actually, it was realised a long time ago
\cite{HTM77,Cheng:1980qt,SchValle80} that a Majorana mass term for the
left-handed (LH) flavour neutrino fields can be generated by
$SU(2)_L\times U(1)_Y$ invariant Yukawa couplings of two lepton
doublet fields to a Higgs triplet field, carrying two units of the
weak hyperchange, $|Y| = 2$.  Such a Higgs field has an electrically
neutral, singly charged and doubly charged components. The Majorana
mass term for the active flavour neutrinos arises when the neutral
component of the Higgs triplet field acquires a nonzero vacuum
expectation value (vev), breaking the $SU(2)_L\times U(1)_Y$ symmetry.
There are several possible realisations of this scenario.  The
realisation in which the global $U(1)_L$ symmetry associated with the
conservation of the total lepton charge $L$ is broken only
spontaneously by the Higgs triplet vev \cite{Gelmini:1980re}, was
ruled out by the LEP data on the invisible decay width of the
$Z^0$-boson.  If, however, the $U(1)_L$ symmetry is broken explicitly
in a manner that leads to a nonzero vacuum expectation value of the
neutral component of the Higgs triplet field (see, e.g.
\cite{Ma:2000wp,Chun:2003ej}) one obtains a viable model of neutrino
mass generation. This model has been investigated in detail recently
\cite{Akeroyd:2007zv,Garayoa:2007fw,Kadastik:2007yd} and was shown to
have a rich and physically interesting phenomenology owing to the fact
that i) the couplings of the doubly and singly charged Higgs fields to
the flavour neutrinos and charged leptons are proportional to the
elements of the Majorana mass matrix of the (flavour) neutrinos,
$M_{{l'}l}$, and can be relatively large, and that ii) the physical
doubly charged and singly charged Higgs fields, $H^{\pm\pm}$ and
$H^{\pm}$, can have masses in the range from $\sim$100 GeV to $\sim$1
TeV and thus can, in principle, be produced and observed at LHC.
Point i) implies that the indicated couplings are determined
essentially by the elements of the Pontecorvo-Maki-Nakagawa-Sakata
(PMNS) neutrino mixing matrix \cite{BPont57} and by the neutrino
masses. In \cite{Akeroyd:2007zv,Garayoa:2007fw,Kadastik:2007yd} it was
shown that by studying the decays $H^{\pm\pm}\to l^\pm {l'}^\pm$,
$l,{l'} = e,\mu,\tau$, it might be possible to obtain information on
the absolute neutrino mass scale, on the type of neutrino mass
spectrum (which can be, e.g. normal hierarchical (NH), inverted
hierarchical (IH) and quasi-degenerate (QD)), and on the Majorana CP
violating phases \cite{BHP80} present in the neutrino mixing matrix.

  In the present article we investigate the possibility to use the
information on the neutrino mass spectrum and the Majorana CP
violating phases from the measurements of the $H^{\pm\pm}\to l^\pm{l'}^\pm$ 
decay branching ratios, $\BR(H^{\pm\pm}\to l^\pm {l'}^\pm)$,
$l,l'=e,\mu$, in order to obtain predictions for the effective
Majorana mass in neutrinoless double beta ($\betabeta$-) decay,
$|M_{ee}|$ (see, e.g. \cite{BiPet87}).  Our study is motivated by the
fact that most probably the searches for the doubly charged scalars
$H^{\pm\pm}$ and the decays $H^{\pm\pm}\to l^\pm {l'}^\pm$ will be
carried out at LHC before the next generation of $\betabeta$-decay
experiments will be operative. Among the different decay channels
$H^{\pm\pm}\to l^\pm {l'}^\pm$, $l,{l'} = e,\mu,\tau$, the easier to
observe are those with two electrons (positrons), two muons
(antimuons), or an electron (positron) and a muon (antimuon),
$e^{\pm}e^{\pm}$, $\mu^{\pm}\mu^{\pm}$ and $e^{\pm}\mu^{\pm}$, in the
final state. If the mass of $H^{\pm\pm}$ does not exceed approximately
400 GeV, the branching ratios of the $H^{\pm\pm}$ decays into
$e^{\pm}e^{\pm}$, $e^{\pm}\mu^{\pm}$ and $\mu^{\pm}\mu^{\pm}$ can be
measured at LHC with a few percent error \cite{Perez:2008ha}.  We will
show that if the doubly charged Higgs bosons $H^{\pm\pm}$ will be
discovered at LHC and at least the three branching ratios
$\BR(H^{\pm\pm}\to e^\pm e^\pm)$, $\BR(H^{\pm\pm}\to e^\pm \mu^\pm)$
and $\BR(H^{\pm\pm}\to \mu^\pm \mu^\pm)$ will be measured with a
sufficient accuracy, one can obtain unique information on the
$\betabeta$-decay effective Majorana mass $|M_{ee}|$.  This
information will be extremely important, in particular, for the
upcoming next generation of $\betabeta$-decay experiments.

%%%%%%%%%%%%%%%%%%%%%%%%%%%%%%%%%%%
\section{The Higgs Triplet Model}
%
%%%%%%%%%%%%%%%%%%%%%%%%%%%%%%%%%%%%
%

In the Higgs Triplet Model~(HTM) \cite{HTM77,Cheng:1980qt,SchValle80}
a $I=1,Y=2$ complex $SU(2)_L$ triplet of Higgs scalar fields is added
to the Standard Model (SM) Lagrangian.  In the $2\times 2$
representation, the Higgs triplet field has the form:
%%%%%%%%%%%%%%%%%%%%%%%%%%%%%%%%%%%
\begin{equation}
\Delta
=\left( \begin{array}{cc}
\Delta^+/\sqrt{2}  & \Delta^{++} \\
\Delta^0       & -\Delta^+/\sqrt{2}
\end{array}\right)
\end{equation}
%%%%%%%%%%%%%%%%%%%%%%%%%%%%%%%%
%
where $\Delta^0$, $\Delta^+$ and $\Delta^{++}$ are neutral, singly
charged and doubly charged scalar fields. In the flavour basis in
which the charged lepton mass matrix is diagonal we are going to use
throughout this article, a Majorana mass term for the LH flavour
neutrinos can be generated (without the introduction of a right-handed
neutrino fields) by the $SU(2)_L\times U(1)_Y$ gauge invariant Yukawa
interaction:
%%%%%%%%%%%%%%%%%%%%%%%%%%%%%%%%%
\begin{equation}
{\cal L} = h_{{l'}l}\,\psi_{{l'}L}^T\,C\,(i\tau_2)\,\Delta\,\psi_{lL} + h.c.
\label{trip_yuk}
\end{equation}
%%%%%%%%%%%%%%%%%%%%%%%%%%%%%%%%%%%
%
Here $h_{{l'}l}=h_{l{l'}}$, ${l'},l=e,\mu,\tau$, are complex Yukawa
couplings forming a symmetric matrix ${\bf h}$, $C$ is the charge
conjugation matrix, $\tau_2$ is a Pauli matrix for $SU(2)_L$ indices,
and $\psi^T_{lL}=(\nu_{lL}~~l_L)^T$, $l=e,\mu,\tau$, is the LH lepton
doublet field.  A non-zero triplet vacuum expectation value,
$\langle\Delta^0\rangle \equiv v_{\Delta} \neq 0$, gives rise to a
Majorana mass matrix ${\bf M}$ for the LH flavour neutrino fields
$\nu_{lL}$:
%%%%%%%%%%%%%%%%%%%%%%%%%%%%%%%
\begin{equation}
M_{{l'}l}=2\,h_{{l'}l}\,\langle\Delta^0\rangle = \sqrt{2}\,h_{{l'}l}\,v_{\Delta}\,,
~~~{l'},l=e,\mu,\tau\,.
\label{nu_mass}
\end{equation}
%%%%%%%%%%%%%%%%%%%%%%%%%%%%%%%
%
The requisite $v_{\Delta}\neq 0$ arises from the minimisation of
the most general $SU(2)\times U(1)_Y$ invariant Higgs potential 
\cite{Ma:2000wp, Chun:2003ej}:
%%%%%%%%%%%%%%%%%%%%%%%%%%%%%%%%%%%%
\begin{eqnarray}
V&=&m^2\,(\Phi^\dagger\Phi) + \lambda_1\,(\Phi^\dagger\Phi)^2\, +
M_{\Delta}^2\, {\rm Tr}(\Delta^\dagger\Delta) +
\lambda_2\,[{\rm Tr}(\Delta^\dagger\Delta)]^2 +
\lambda_3\, {\rm Det}(\Delta^\dagger\Delta)  \nonumber \\
&& + \lambda_4\,(\Phi^\dagger\Phi){\rm Tr}(\Delta^\dagger\Delta)
+ \lambda_5\,(\Phi^\dagger\tau_i\Phi){\rm Tr}(\Delta^\dagger\tau_i
\Delta) + \left(
{1\over \sqrt 2}\mu\,(\Phi^Ti\tau_2\Delta^\dagger\Phi) + h.c \right)\,,
\label{higgs_potential}
\end{eqnarray}
%%%%%%%%%%%%%%%%%%%%%%%%%%%%%%%%%%%%
%
$\Phi^T=(\phi^+~~\phi^0)^T$ being the SM Higgs doublet field. In 
eq.~(\ref{higgs_potential}), $M_{\Delta}^2\,> 0$ is the common mass of the
triplet scalars.  The choice $m^2 < 0$ ensures that
$\langle\phi^0\rangle=v/\sqrt 2 \neq 0$, which breaks spontaneously
$SU(2)\times U(1)_Y$ to $U(1)_Q$.  In the model of
Gelmini-Roncadelli~\cite{Gelmini:1980re} the term
$\mu(\Phi^Ti\tau_2\Delta^\dagger\Phi)$ is absent, which leads for
$M_{\Delta}^2<0$ to a spontaneous breaking of the global $U(1)_L$
symmetry associated with the conservation of the total lepton number.
The resulting Higgs spectrum contains a massless Goldstone boson - the
triplet scalar Majoron, $J$, and another light scalar, $H^0$.  The
decay $Z^0 \to H^0J$ would give too large a contribution to the
invisible decay width of the $Z^0$-boson and this model was excluded
by the LEP data.  The inclusion of the term
$\mu(\Phi^Ti\tau_2\Delta^\dagger\Phi$) explicitly breaks the lepton
number conservation when $\Delta$ is assigned two units of the total
lepton charge $L$, and therefore avoids the presence of a Goldstone
boson - the Majoron, in the model~\cite{Cheng:1980qt,SchValle80}.
Thus, the scalar potential in eq.~(\ref{higgs_potential}) together
with the triplet Yukawa interaction of eq.~(\ref{trip_yuk}) lead to a
phenomenologically viable model of neutrino mass generation.

 The expression for $v_\Delta$ resulting from the minimisation of the
potential $V$, eq.~(\ref{higgs_potential}), reads:
%%%%%%%%%%%%%%%%%%%%%%%%%%%%%%%%%%%%%%%%%%
\begin{equation}
v_\Delta \simeq \frac{\mu v^2}
{2M_{\Delta}^2 +(\lambda_4 + \lambda_5)\,v^2}\,,~~~
{\rm for}~v_\Delta \ll v\,.
\label{triplet_vev}
\end{equation}
%%%%%%%%%%%%%%%%%%%%%%%%%%%%%%%%%%%%%%%%%%
%
 In the scenario of relatively light triplet scalars within the
discovery reach of the LHC we will be interested in, one has
$M_{\Delta}\approx v$ and eq.~(\ref{triplet_vev}) leads to
$v_\Delta\approx \mu$.  In extensions of the HTM, the term
$\mu(\Phi^Ti\tau_2\Delta^\dagger\Phi$) can arise in various ways: i)
through the vev of a Higgs singlet field~\cite{Schechter:1981cv, Diaz:1998zg}; 
ii) can be generated at higher
orders in perturbation theory~\cite{Chun:2003ej}; or iii) can appear
in the context of theories with extra dimensions~\cite{Ma:2000wp, Chen:2005mz}.

 An upper limit on $v_\Delta$ can be obtained from
considering its effect on the parameter $\rho=M^2_W/M_Z^2\cos^2\theta_W$.
In the SM, $\rho=1$ at tree-level, while in the HTM one has
%%%%%%%%%%%%%%%%%%%%%%%%%%%%%%%%%%%
\begin{equation}
\rho\equiv 1+\delta\rho={1+2x^2\over 1+4x^2}\,,~~~x \equiv v_\Delta/v.
\label{deltarho}
\end{equation}
%%%%%%%%%%%%%%%%%%%%%%%%%%%%%%%%%%%
%
The measurement $\rho\approx 1$ leads to the bound
$v_\Delta/v\lesssim 0.03$, or  $v_\Delta<8$~GeV.
At the 1-loop level $v_\Delta$ must be renormalised and explicit
analyses lead to bounds on its magnitude similar to those derived from
the tree-level analysis~\cite{tree}.

  The HTM has seven physical Higgs scalar particles
$(H^{++},H^{--},H^+,H^-,H^0,A^0,h^0)$.  The doubly charged Higgs field
$H^{++}$ coincides with the triplet scalar field $\Delta^{++}$.  The
remaining Higss mass-eigenstate fields are in general mixtures of the
doublet and triplet fields.  The corresponding mixing parameter is
proportional to the ratio of triplet and doublet vevs, $v_\Delta/v$,
and hence is small {\it even if} $v_\Delta$ assumes its largest value
of a few GeV.  Therefore $H^+,H^0,A^0$ are predominantly composed of
the triplet fields, while $h^0$ is predominantly composed of the
doublet field and plays the role of the SM Higgs boson.  The masses of
$H^{\pm\pm},H^\pm,H^0,A^0$ are of order $M_{\Delta}$ with splittings
of order $\lambda_5~v$.  For $M_\Delta < 1$~TeV of interest for direct
searches for the Higgs bosons at the LHC, the couplings $h_{{l'}l}$
are constrained to be ${\cal O}(0.1)$ or less by a variety of
processes such as $\mu\to eee, \tau\to lll$ etc. These constraints are
reviewed in \cite{Gunion:1989in, Cuypers:1996ia}.

  In this article we will be interested, in particular, in the decays of
$H^{\pm\pm}$ into a pair of same-sign charged leptons,
$H^{\pm\pm}\rightarrow {l'}^{\pm}l^{\pm}$, ${l'},l=e,\mu,\tau$, which
give clear signals even in hadron colliders like the LHC.  These
decays are important not only for the searches of $H^{\pm\pm}$, but
also because of the very interesting possibility that their lepton
flavour dependence can be directly related to the Majorana mass matrix
of the LH flavour neutrinos.

   In our analysis we will assume that $M_{H^{\pm\pm}} \leq M_{H^\pm}$
and $v_\Delta \lesssim 1~\text{MeV}$. Under these conditions
the decay $H^{\pm\pm}\to H^\pm W^\pm$ is forbidden,
while the decay $H^{\pm\pm}\to W^\pm W^\pm$ is
sufficiently strongly suppressed. In this case the
branching ratios of the decays $H^{\pm\pm}\rightarrow {l'}^{\pm}l^{\pm}$
are given by the following simple expressions
 (see, e.g. \cite{Akeroyd:2007zv,Garayoa:2007fw,Kadastik:2007yd}):
%%%%%%%%%%%%%%%%%%%%%%%%%%%%%%%%%%%%%%%%%%%%%%
\begin{eqnarray}
\BR_{{l'}l}
 \equiv \BR(H^{\pm\pm}\to {l'}^\pm l^\pm)
&=&
 \frac{2}{1+\delta_{{l'}l}}\,
 \frac{|h_{{l'}l}|^2}{\sum_{{l'}l} |h_{{l'}l}|^2}\\
&=&
 \frac{2}{1+\delta_{{l'}l}}\,
 \frac{ \left| M_{{l'}l} \right|^2 }{ \sum_i m_i^2 },
\end{eqnarray}
%%%%%%%%%%%%%%%%%%%%%%%%%%%%%%%%%%%%%%%%%%%%%
%
where $\delta_{{l'}l}$ is the Kronecker delta.  Note that the
branching ratios depend only on the parameters of neutrino mass
matrix.  The measurement of $\BR_{{l'}l}$ can give significant
information on the elements of the neutrino mass matrix 
$|M_{{l'}l}|$, and therefore, e.g. on the absolute neutrino mass scale
(i.e. lightest neutrino mass), type of neutrino mass spectrum,
Majorana CP violating phases in the neutrino mixing matrix, etc.

%%%%%%%%%%%%%%%%%%%%%%%%%%%%%%%%%%%%%%%%%%%%%%
%
\section{The Neutrino Masses, Mixing and the $\betabeta$-Decay}
%
%%%%%%%%%%%%%%%%%%%%%%%%%%%%%%%%%%%%%%%%%%%%%%
%
  We work in the flavour basis in which the mass matrix of the charged
leptons is diagonal.  As we have shown, in the Higgs triplet model of
interest, the LH flavour neutrino fields $\nu_{lL}$ acquire a Majorana
mass term.  The corresponding Majorana mass matrix ${\bf M}$ is
diagonalised with the help of the Pontecorvo-Maki-Nakagawa-Sakata
(PMNS) neutrino mixing matrix \cite{BPont57}:
%%%%%%%%%%%%%%%%%%%%%%%%%%%%%%%%
\begin{eqnarray}
M_{l{l'}}
 = [U_\PMNS\, \text{diag}(m_1, m_2, m_3)\, U_\PMNS^T]_{l{l'}},
\end{eqnarray}
%%%%%%%%%%%%%%%%%%%%%%%%%%%%%%%%
%
where $m_j$, $j=1,2,3$, are the real positive eigenvalues of
$M_{l{l'}}$ - the masses of the Majorana neutrinos $\chi_j$ with
definite mass.  In what follows we will use the standard
parametrisation of the PMNS matrix (see, $\eg$ \cite{BPP1,STPNu04}):
%%%%%%%%%%%%%%%%%%%%%%%%%%%%%%%%%%%
\begin{equation}
\begin{array}{c}
\label{eq:Upara}
 U_{\rm PMNS} \equiv \left( \begin{array}{ccc}
 c_{12} c_{13} & s_{12} c_{13} & s_{13} e^{-i \delta} \\[0.2cm]
 -s_{12} c_{23} - c_{12} s_{23} s_{13} e^{i \delta}
 & c_{12} c_{23} - s_{12} s_{23} s_{13} e^{i \delta} & s_{23} c_{13} \\[0.2cm]
 s_{12} s_{23} - c_{12} c_{23} s_{13} e^{i \delta} &
 - c_{12} s_{23} - s_{12} c_{23} s_{13} e^{i \delta} & c_{23} c_{13} \\
     \end{array}   \right)
{\rm diag}(1, e^{i \frac{\alpha_{21}}{2}}, e^{i \frac{\alpha_{31}}{2}}) \,,
 \end{array} \end{equation}
%%%%%%%%%%%%%%%%%%%%%%%%%%%%%%%%%
%
\noindent where $c_{ij} \equiv \cos\theta_{ij}$, $s_{ij} \equiv
\sin\theta_{ij}$, the angles $\theta_{ij} = [0,\pi/2]$ $(i<j=1,2,3)$,
$\delta = [0, 2\pi]$ is the Dirac CP-violating phase, and $\alpha_{21}$
and $\alpha_{31}$ are two Majorana CP-violation
phases~\cite{BHP80,Doi81}. The phases $\alpha_{21}$ and $\alpha_{31}$
can vary in the interval $[0, 2\pi]$. It proves useful for our further
discussion to define also the difference of the two Majorana phases:
$\alpha_{32}\equiv \alpha_{31}-\alpha_{21}$.  Let us add that at
present we do not have experimental information on $\delta$,
$\alpha_{21}$ and $\alpha_{31}$.

The existing neutrino oscillation data~\cite{solar,atm,acc,Apollonio:2002gd,:2008ee} 
allow to determine with
a rather good precision the mixing angles and neutrino mass squared
differences which drive the solar neutrino and the dominant
atmospheric neutrino oscillations, $\sin^2{2\theta_{12}}$, 
$\Delta m^2_{21}$ and $\sin^2{2\theta_{23}}$, 
$|\Delta m^2_{31}| (\cong|\Delta m^2_{32}|)$, and to obtain a rather stringent limit on the
CHOOZ angle $\theta_{13}$.  In our analysis we will use the following
best fit values of $\sin^2{2\theta_{12}}$, $\Delta m^2_{21}$,
$\sin^2{2\theta_{23}}$ and $|\Delta m^2_{31}|$ 
\cite{BCGPRKL2,Fogli08,TSchw08}:
%%%%%%%%%%%%%%%%%%%%%%%
\begin{eqnarray}
\label{eq:solvalues}
\Delta m^2_{21} = 7.6 \times 10^{-5} \ \eV^2,  &&
\sin^2{2\theta_{12}} = 0.87, \\
\label{eq:atmvalues}
|\Delta m^2_{31}| = 2.4 \times 10^{-3} \ \eV^2, &&
\sin^2{2\theta_{23}} = 1\,.% \\% \sin^2{2\theta_{23}} > 0.94, \\
\end{eqnarray}
%%%%%%%%%%%%%%%%%%%%%%%%%%%%%%%%%%%%%%%%%%%%%
%
The upper limit on $\sin^2{2\theta_{13}}$ obtained
in CHOOZ reactor anti-neutrino experiment~\cite{Apollonio:2002gd} 
reads:
%%%%%%%%%%%%%%%%%%%%%%%%%%%%%%%%
\begin{eqnarray}
\sin^2{2\theta_{13}} < 0.14\,.
\end{eqnarray}
%%%%%%%%%%%%%%%%%%%%%%%%%%%%%%%%%%%%%%%%%%%%%
%
From the global analyses of the neutrino oscillation data one finds
(see, e.g. \cite{TSchw08}):
%%%%%%%%%%%%%%%%%%%%%%%%%%%%%%%%
\begin{eqnarray}
\sin^2{\theta_{13}} < 0.056\,,~~~99.73\%~{\rm C.L.}
\end{eqnarray}
%%%%%%%%%%%%%%%%%%%%%%%%%%%%%%%%%%%%%%%%%%%%%
%
The next generation of experiments with reactor $\overline{\nu}_e$,
which are under preparation, Dooble CHOOZ \cite{DCHOOZ}, Daya Bay
\cite{DayaB}, RENO \cite{RENO}, can improve the currently reached
sensitivity to the value of $\sin^2{2\theta_{13}}$ by a factor of
(5-10) (see, $\eg$~\cite{Reacth13}), while future long baseline
experiments aim at measuring values of $\sin^2{2\theta_{13}}$ as small
as $10^{-4}$-$10^{-3}$ (see, e.g.~\cite{Bandyopadhyay:2007kx}).

 Let us note that the uncertainty in the experimental determination of
$\sin^2{2\theta_{23}}$ corresponds to a rather large interval of
allowed values of $s^2_{23}$ \cite{atm}: $0.38 \leq s_{23}^2 \leq
0.62$.  We will take into account this uncertainty in our numerical
analysis \footnote{Varying $\sin^2\theta_{12}$ in the $3\sigma$
  interval of allowed values of $\sin^2\theta_{12}$
  \cite{BCGPRKL2,TSchw08}, $0.25 \ltap \sin^2\theta_{12} \ltap 0.37$,
  has essentially negligible effect on the results of our analysis.}.
It should be added that the accuracy on $\sin^2{2\theta_{23}}$ is
planned to be improved considerably in future long baseline
experiments. The uncertainty in $\sin^2{2\theta_{23}}$ is foreseen to
be reduced in the T2K experiment~\cite{T2K}, for instance, to
$\sin^2{2\theta_{23}} > 0.99$ ($0.45 < s_{23}^2 < 0.55$), if the true
value of $\sin^2{2\theta_{23}} = 1$.  As we will see, the correlations
between the branching ratios of the decays $H^{\pm\pm}\to l^\pm {l'}^\pm$, 
$\BR_{l{l'}}$, $l,{l'} = e,\mu,\tau$, and the effective
Majorana mass in neutrinoless double beta ($\betabeta$-) decay, 
$\meff\equiv M_{ee}$, which is the main subject of our study, depend not
only on the elements of the neutrino mixing matrix, but also on the
type of neutrino mass spectrum and on the absolute scale of neutrino
masses.

  As is well known, owing to the fact that the sign of 
$\Delta m^2_{31}$, cannot be determined from the existing data, there are two
possible types of neutrino mass spectrum compatible with the data -
with normal ordering and with inverted ordering.  In the standardly
used convention we are also going to employ, the two spectra
correspond to:
%%%%%%%%%%%%%%%%%%%%%%%%%%%
\begin{flushleft}
\begin{itemize}
\item[--] $m_1 < m_2 < m_3$, $\Delta m^2_{31} >0$,
  {\it normal ordering (NO)},
\item[--] $m_3 < m_1 < m_2$, $\Delta m^2_{31}< 0$,
{\it inverted ordering (IO)}.
\end{itemize}
\end{flushleft}
%%%%%%%%%%%%%%%%%%%%%%%%%%%%
%
The $\nu$-mass spectrum can be: i) {\it normal hierarchical (NH)},
$m_1 \ll m_2 < m_3$, with $m_2 \cong \sqrt{\Delta m^2_{21}} \cong
8.8\times 10^{-3}$ eV, $m_3 \cong \sqrt{\Delta m^2_{31}} \cong
4.9\times 10^{-2}$ eV; ii) {\it inverted hierarchical (IH)}, $m_3 \ll
m_1 < m_2$, with $m_2 \cong \sqrt{\Delta m^2_{23}} \cong 4.9\times
10^{-2}$ eV; $m_1 \cong \sqrt{\Delta m^2_{23} - \Delta m^2_{21}} \cong
4.8\times 10^{-2}$ eV, and iii) {\it quasi-degenerate (QD)}, $m_1
\cong m_2 \cong m_3$, $m^2_{1,2,3} \gg |\Delta m^2_{31}|$.  In the
latter case one has: $m_{j}\gtap 0.10$ eV.

 The type of neutrino mass spectrum, i.e. 
${\rm sgn}(\Delta m^2_{31})$, can be determined by studying oscillations 
of neutrinos
and antineutrinos, say, $\nu_{\mu} \leftrightarrow \nu_e$ and
$\bar{\nu}_{\mu} \leftrightarrow \bar{\nu}_e$, in which matter effects
are sufficiently large.  This can be done in long base-line
$\nu$-oscillation experiments (see, e.g.\
\cite{Bandyopadhyay:2007kx}).  If $\sin^22\theta_{13}\gtap 0.05$ and
$\sin^2\theta_{23}\gtap 0.50$, information on ${\rm sgn}(\Delta
m^2_{31})$ might be obtained in atmospheric neutrino experiments by
investigating the effects of the subdominant transitions $\nu_{\mu(e)}
\rightarrow \nu_{e(\mu)}$ and $\bar{\nu}_{\mu(e)} \rightarrow
\bar{\nu}_{e(\mu)}$ of atmospheric neutrinos which traverse the Earth
\cite{JBSP203}.  For $\nu_{\mu(e)}$ ({\it or} $\bar{\nu}_{\mu(e)}$)
crossing the Earth core, new type of resonance-like enhancement of the
indicated transitions takes place due to the {\it (Earth) mantle-core
  constructive interference effect (neutrino oscillation length
  resonance (NOLR))} \cite{SP3198}~\footnote{As a consequence of this
  effect, the corresponding $\nu_{\mu(e)}$ ({\it or}
  $\bar{\nu}_{\mu(e)}$) transition probabilities can be maximal
  \cite{106107} (for the precise conditions of the mantle-core (NOLR)
  enhancement see \cite{SP3198,106107}).  Let us note that the Earth
  mantle-core (NOLR) enhancement of neutrino transitions differs
  \cite{SP3198} from the MSW one. The conditions of the Earth
  mantle-core enhancement \cite{SP3198,106107} also differ
  \cite{PRL85ChP} from the conditions of the parametric resonance
  enhancement of the neutrino transitions discussed in the articles
  \cite{Param86}.}.  For $\Delta m^2_{31}> 0$, the neutrino
transitions $\nu_{\mu(e)} \rightarrow \nu_{e(\mu)}$ are enhanced,
while for $\Delta m^2_{31}< 0$ the enhancement of antineutrino
transitions $\bar{\nu}_{\mu(e)} \rightarrow \bar{\nu}_{e(\mu)}$ takes
place, which might allow to determine ${\rm sgn}(\Delta m^2_{31})$.
If $\sin^2\theta_{13}$ is sufficiently large, the sign of $\Delta
m^2_{31}$ can also be determined by studying the oscillations of
reactor $\bar{\nu}_e$ on distances of $\sim (20 -40)$ km
\cite{PiaiP0103}.  An experiment with reactor $\bar{\nu}_e$, which, in
particular, might have the capabilities to measure ${\rm sgn}(\Delta
m^2_{31})$, was proposed recently in \cite{Hano} (see also \cite{DBhier08}).
% According to \cite{Hano} (see also \cite{DBhier08}),
% this experiment can provide
% a determination of $|\deltaatm|$
% with an uncertainty of
% $(3 - 4)\%$ at 3$\sigma$.
Information on the type of neutrino mass spectrum
can also be obtained in $\beta$-decay experiments
having a sensitivity to neutrino masses 
$\sim \sqrt{|\deltaatm|}\cong 5\times 10^{-2}$ eV~\cite{BMP06}
(i.e.\ by a factor of $\sim 4$ better than the
planned sensitivity of the  KATRIN experiment~\cite{MainzKATRIN}, 
see below).

Direct information on the absolute neutrino mass scale can be derived
in \hbeta~experiments~\cite{Fermi34,MoscowH3,MainzKATRIN}.  The most stringent upper bounds
on the $\bar{\nu}_e$ mass were obtained in the Troitzk~\cite{MoscowH3}
and Mainz~\cite{MainzKATRIN} experiments: \vspace{-0.10cm}
%%%%%%%%%%%%%%%%%%%%%%%%%%%%%%%%%%%%%%%
\begin{equation}
m_{\bar{\nu}_e} < 2.3~{\rm eV},~~~95\%~{\rm C.L.}
\label{H3beta}
\end{equation}
%%%%%%%%%%%%%%%%%%%%%%%%%%%%%%
%
\noindent We have $m_{\bar{\nu}_e} \cong m_{1,2,3}$
in the case of the QD $\nu$-mass spectrum.
The KATRIN experiment~\cite{MainzKATRIN},
which is under preparation,
is planned to reach a sensitivity
of  $m_{\bar{\nu}_e} \sim 0.20$~eV,
i.e. it will probe the region of the QD
spectrum.

The CMB data of the WMAP experiment \cite{WMAPnu}, combined with data
from large scale structure surveys (2dFGRS, SDSS), lead to the
following upper limit on the sum of neutrino masses (see, e.g. \cite{Hann06}):
%%%%%%%%%%%%%%%%%%%%%%%%%%%%%%%%%%%%%%%
\begin{equation}
\sum_{j} m_{j} \equiv \Sigma < (0.4 \mbox{--} 1.7)~
{\rm eV}\,,~~~95\%~{\rm C.L.}
\label{WMAP}
\end{equation}
%%%%%%%%%%%%%%%%%%%%%%%%%%%%%%
%
Data on weak lensing of galaxies,
combined with data from the WMAP and PLANCK
experiments, may allow $\Sigma$ to be determined
with an uncertainty of $\sim 0.04$~eV~\cite{Hann06,Hu99}.

  In our analysis we will consider both types of
neutrino mass spectrum - with normal and with inverted
ordering, as well as the specific cases of normal hierarchical (NH),
inverted hierarchical (IH) and quasi-degenerate (QD) spectra.
Correspondingly, the lightest neutrino mass ${\rm min}(m_j) \equiv m_0$,
which determines the absolute neutrino mass scale,
will be varied in the interval:
%%%%%%%%%%%%%%%%%%%%%%%%%%%%%%%%%%%%%%%
\begin{equation}
0~\leq~ m_0~ \leq~ 0.3~{\rm eV}\,,~~~m_0 \equiv {\rm min}(m_j)\,,~j=1,2,3.
\label{minmj}
\end{equation}
%%%%%%%%%%%%%%%%%%%%%%%%%%%%%%
%
As we will show, the results we obtain essentially do not depend on
the maximal value $m_0$ as long as the latter is not smaller than
$0.3$ eV. The reason is that for $m_0~\gtap~ 0.3~{\rm eV}$ (i.e. in
the QD region), the branching ratios $\BR_{{l'}l}$ we are interested
in practically do not depend on the neutrino masses:
%%%%%%%%%%%%%%%%%%%%%%%%%%%%%%%%%%%%%%%
\begin{equation}
\BR_{{l'}l} \cong \frac{2}{3(1+\delta_{{l'}l})}\,
 \left |\sum_{j} U_{{l'}j}U_{lj} \right|^2\,,~~~m_0~\gtap~ 0.3~{\rm eV}\,.
\label{satBl{l'}}
\end{equation}
%%%%%%%%%%%%%%%%%%%%%%%%%%%%%%
%

%%%%%%%%%%%%%%%%%%%%%%%%%%%%%%%%%%%%%%%%%%%%%%%
%
\leftline{\bf Neutrinoless Double Beta decay}
%
%%%%%%%%%%%%%%%%%%%%%%%%%%%%%%%%%%%%%%%%%%%%
%

In the Higgs triplet model the massive neutrinos are Majorana
particles.  Determining the nature of massive neutrinos is of
fundamental importance for understanding the origin of neutrino masses
and, more generally, for understanding the symmetries governing the
particle interactions.  The existence of massive Majorana neutrinos is
associated with non-conservation of the total lepton charge.  In this
case the neutrinoless double beta decay $(A,Z) \to (A, Z+2) + 2e^-$ is
allowed (see, e.g. \cite{BiPet87, APSbb0nu, STPFocusNu04}).
%%%%%%%%%%%%%%%%%%%%%%%%%%%%%%%%%%%%%%%%%
\begin{figure}
\begin{center}
\hspace*{0mm}(a)\hspace*{25mm}(b)\hspace*{25mm}(c)\\[1mm]
\includegraphics[angle=0,width=3cm]{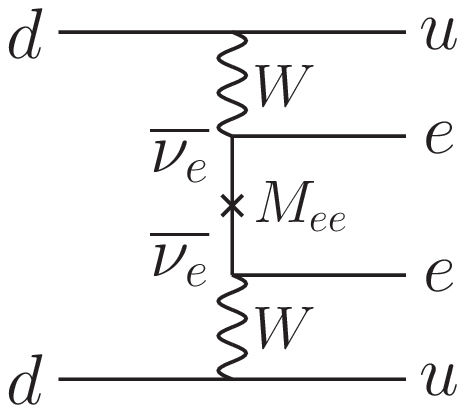}
\includegraphics[angle=0,width=3cm]{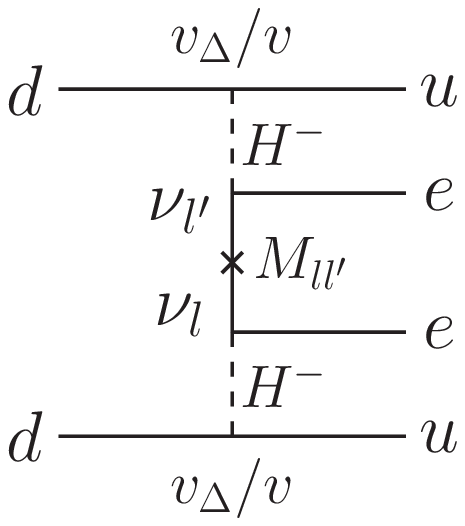}
\includegraphics[angle=0,width=3cm]{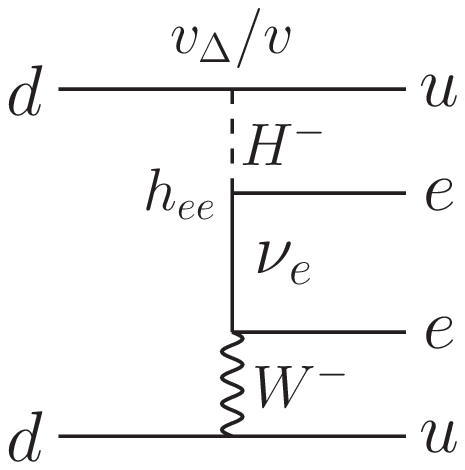}\\[3mm]
\hspace*{0mm}(d)\hspace*{25mm}(e)\hspace*{25mm}(f)\\[2mm]
\includegraphics[angle=0,width=3cm]{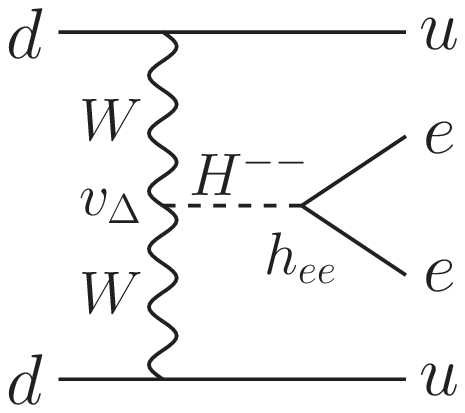}
\includegraphics[angle=0,width=3cm]{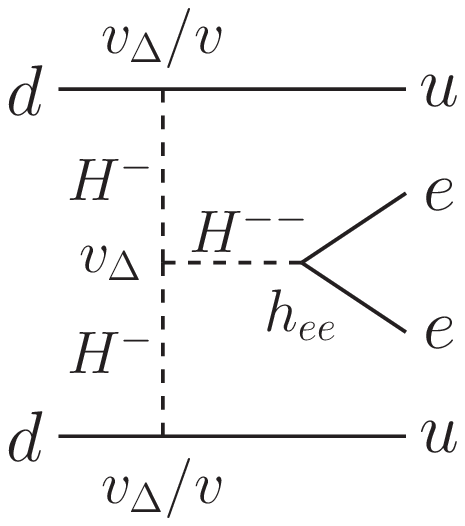}
\includegraphics[angle=0,width=3cm]{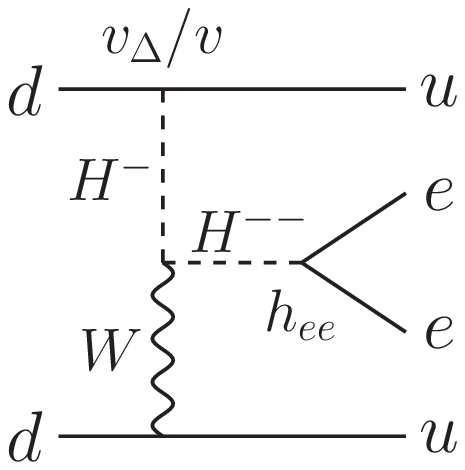}
\caption{Diagrams for the neutrinoless double beta decay.
The diagram (a) is the standard and dominant one,
and (b)-(f) are possible but negligible in the HTM\@.}
\label{fig:diag-dbeta}
\end{center}
\end{figure}
%%%%%%%%%%%%%%%%%%%%%%%%%%%%%%%%%%%%%%%%%%%%%%%%
%
Assuming that the dominant mechanism for the decay
is the exchange of light Majorana neutrinos
(Fig.~\ref{fig:diag-dbeta}(a)),
the half-life $T_{1/2}^{0\nu}$ for the decay is given by
%%%%%%%%%%%%%%%%%%%%%%%%%%%%%%%%%%%%%%%%%%%
\begin{eqnarray}
 T_{1/2}^{0\nu}
 = \left( G^{0\nu} |M^{0\nu}|^2 |M_{ee}|^2 \right)^{-1}\,,
\end{eqnarray}
%%%%%%%%%%%%%%%%%%%%%%%
%
where $G^{0\nu}$ is a phase space factor and $M^{0\nu}$ is the nuclear
matrix element of the process.  All the dependence of $T_{1/2}^{0\nu}$
on the neutrino masses and mixing parameters factorises into the
$\betabeta$-decay effective Majorana mass $M_{ee}$:
%%%%%%%%%%%%%%%%%%%%%%%%%%%%%%%%%%%%%
\begin{eqnarray}
 |M_{ee}| =  \left| c_{12}^2 c_{13}^2 m_1
   + s_{12}^2 c_{13}^2 m_2 e^{i\alpha_{21}}
   + s_{13}^2 m_3 e^{i(\alpha_{31}-2\delta)} \right|\,.
\end{eqnarray}
%%%%%%%%%%%%%%%%%%%%%%%%%%%%%%%%%%%%
%

The most stringent upper bound on $|M_{ee}|$, $|M_{ee}| <
(0.35\,\text{-}\,1.05)$ eV was obtained by using the lower limit
$T_{1/2}^{0\nu} > 1.9\times 10^{25}~\text{yr}$ (90\%~C.L.) found
\footnote{In the quoted upper bound for $|M_{ee}|$ a factor of 3
  uncertainty in the relevant NME (see, e.g.\ \cite{FesSimVogel03}) is
  taken into account.}  in the Heidelberg-Moscow $^{76}$Ge
experiment~\cite{HMGe76}.  The IGEX collaboration has obtained the
result $T_{1/2}^{0\nu} > 1.57\times 10^{25}~\text{yr}$ (90\%~C.L.),
from which the limit $|M_{ee}| < (0.33 - 1.35)$~eV was derived
\cite{IGEX00}.  A positive $\betabeta$-decay signal at $> 3\sigma$,
corresponding to $T_{1/2}^{0\nu} = (0.69 - 4.18)\times
10^{25}~\text{yr}$ (99.73\%~C.L.) and implying $|M_{ee}|= (0.1 -
0.9)~{\rm eV}$, is claimed to be observed in \cite{Klap04}, while a
recent analysis reports evidence at 6$\sigma$ of $\betabeta$-decay
with $|M_{ee}|= 0.32 \pm 0.03$~eV~\cite{KlapdorMPLA}.  Two
experiments, NEMO3 (with $^{100}$Mo, $^{82}$Se, etc.)~\cite{NEMO3} and
CUORICINO (with $^{130}$Te)~\cite{CUORI}, designed to reach a
sensitivity to $|M_{ee}|\sim (0.2-0.3)$ eV, set the limits: $|M_{ee}|
< (0.61\,\text{-}\,1.26)$~eV~\cite{NEMO3} and $|M_{ee}| < (0.19 -
0.68)$~eV~\cite{CUORI} (90\% C.L.), where estimated uncertainties in
the NME are accounted for. The two upper limits were derived from the
experimental lower limits on the half-lifes of $^{100}$Mo and
$^{130}$Te, $T_{1/2}^{0\nu} > 5.8\times 10^{23}~\text{yr}$ (90\%C.L.)~\cite{NEMO3} 
and $T_{1/2}^{0\nu} > 3.0\times 10^{24}~\text{yr}$
(90\%C.L.)~\cite{CUORI}.  Most importantly, a large number of projects aim at a
sensitivity to $|M_{ee}| \sim (0.01 - 0.05)$ eV \cite{bb0nu}: CUORE
($^{130}$Te), GERDA ($^{76}$Ge), SuperNEMO, EXO ($^{136}$Xe), MAJORANA
($^{76}$Ge), MOON ($^{100}$Mo), COBRA ($^{116}$Cd), XMASS
($^{136}$Xe), CANDLES ($^{48}$Ca), etc.  These experiments, in
particular, will test the positive result claimed in \cite{Klap04}.

The predicted value of $|M_{ee}|$ depends strongly on the type of
$\nu-$mass spectrum \cite{PPSNO2bb,BPP1}, more precisely, on the type
of hierarchy neutrino masses obey.  The existence of significant and
robust lower bounds on $|M_{ee}|$ in the cases of IH and QD 
spectra~\cite{PPSNO2bb} (see also \cite{PPW}), given respectively
\footnote{Up to small corrections we have in the cases of two spectra
  \cite{PPSNO2bb}: $|M_{ee}| \gtap |\Delta m^2_{32}\cos2\theta_{12}|$
  (IH) and $|M_{ee}| \gtap m_0 \cos2\theta_{12}$ (QD).  The
  possibility of $\cos2\theta_{12} = 0$ is ruled out at $\sim 6\sigma$
  by the existing data which also imply that $\cos 2\theta_{12} \gtap
  0.30~(0.26)$ at 2$\sigma$~(3$\sigma$)~\cite{BCGPRKL2,TSchw08}. We
  also have $2.07\times 10^{-3}~{\rm eV^2}\ltap |\Delta m^2_{32}|
  \ltap 2.75\times 10^{-3}~{\rm eV^2}$ at 3$\sigma$.  } by $|M_{ee}|
\gtap 0.01$ eV and $|M_{ee}| \gtap 0.03$~eV, which lie either
partially (IH spectrum) or completely (QD spectrum) within the range
of sensitivity of the next generation of \betabeta-decay experiments,
is one of the most important features of the predictions of
$|M_{ee}|$.  ~At the same time we have $|M_{ee}| \ltap 5\times
10^{-3}$~eV in the case of NH spectrum \cite{PPSchw05}.  The fact that
${\rm max}(|M_{ee}|)$ in the case of NH spectrum is considerably
smaller than ${\rm min}(|M_{ee}|)$ for the IH and QD spectrum opens
the possibility of obtaining information about the type of $\nu$-mass
spectrum from a measurement of $|M_{ee}| \neq 0$~\cite{PPSNO2bb}.
More specifically, a positive result in the future generation of
\betabeta-decay experiments with $|M_{ee}| > 0.01$~eV would imply that
the NH spectrum is strongly disfavoured (if not excluded).  For
$\Delta m^2_{31} > 0$, such a result would mean that the neutrino mass
spectrum is with normal ordering, but is not hierarchical. If $\Delta
m^2_{31} < 0$, the neutrino mass spectrum should be either IH or QD\@.

%%%%%%%%%%%%%%%%%%%%%%%%%%%%%%%%%%%%%%%%%%%%%%%%
%
\section{Prediction for $|M_{ee}|$ from Measurements of
$\BR(H^{\pm\pm}\to {l'}^\pm l^\pm)$}
%
%%%%%%%%%%%%%%%%%%%%%%%%%%%%%%%%%%%%%%%%%%%%%%%
%

In this Section we investigate within the HTM the predictions one can
obtain for the $\betabeta$-decay effective Majorana mass $|M_{ee}|$ by
using data on $\BR_{{l'}l}$.  The dominant mechanism of
$\betabeta$-decay - the light Majorana neutrino exchange, corresponds
to the diagram in Fig.~\ref{fig:diag-dbeta}(a), the contributions from
the diagrams in Figs.~\ref{fig:diag-dbeta}(b)-(f) being 
negligible~\cite{Schechter:1981bd, Mohapatra:1981pm}.

We use three branching ratios, $\BR_{ee}$, $\BR_{e\mu}$, and
$\BR_{\mu\mu}$, in our analysis.  The expressions for these branching
ratios in terms of neutrino masses, neutrino mixing angles and CP
violating phases read (see \cite{Akeroyd:2007zv,Garayoa:2007fw,Kadastik:2007yd}):
%%%%%%%%%%%%%%%%%%%%%%%%%%%%%%%%
\begin{eqnarray}
 ({\textstyle \sum_i} m_i^2) \BR_{ee}&=& |M_{ee}|^2
\nonumber\\
&=&
 \left| c_{12}^2 c_{13}^2 m_1
   + s_{12}^2 c_{13}^2 m_2 e^{i\alpha_{21}}
   + s_{13}^2 m_3 e^{i(\alpha_{31}-2\delta)} \right|^2,
\\
 ({\textstyle \sum_i} m_i^2) \BR_{e\mu}
&=&
 2 \left| c_{12} c_{13}
   ( -s_{12}c_{23} - c_{12}s_{23}s_{13}e^{i\delta})
   m_1 \right.
\nonumber\\
&&\hspace*{5mm}\left.
  {}+ s_{12} c_{13}
     ( c_{12}c_{23} - s_{12}s_{23}s_{13}e^{i\delta})
     m_2 e^{i\alpha_{21}}
    + s_{23}c_{13}s_{13} m_3 e^{i(\alpha_{31}-\delta)} \right|^2,
\\
 ({\textstyle \sum_i} m_i^2) \BR_{\mu\mu}
&=&
 \left| ( -s_{12}c_{23} - c_{12}s_{23}s_{13}e^{i\delta})^2 m_1 \right.
\nonumber\\
&&\hspace*{5mm}\left.  
{}+ ( c_{12}c_{23} - s_{12}s_{23}s_{13}e^{i\delta})^2
     m_2 e^{i\alpha_{21}}
    + s_{23}^2c_{13}^2 m_3 e^{i\alpha_{31}} \right|^2.
\end{eqnarray}
%%%%%%%%%%%%%%%%%%%%%%%%%%%%%%%%%%%%%%%%%%%%%%%
%

Given the solar and atmospheric neutrino oscillation parameters and
the CHOOZ angle, $\sin^2\theta_{12}$, $\Delta m^2_{21}$,
$\sin^2\theta_{23}$, $|\Delta m^2_{31}|$ and $\theta_{13}$, $|M_{ee}|$
depends on $m_0 \equiv {\rm min}(m_j)$, $\alpha_{21}$, $\alpha_{31}$,
$\delta$ and on the type of neutrino mass spectrum (NO or IO). In the
case of spectrum with IO or of QD type, the dependence of $|M_{ee}|$
on $\theta_{13}$ is relatively weak and can be neglected, as long as
\footnote{The inequality $\cos2\theta_{12} \gg \sin^2\theta_{13}$ is
  fulfilled for the 2$\sigma$ experimentally allowed ranges of values
  of $\cos2\theta_{12}$ and $\sin^2\theta_{13}$, see. e.g.
  \cite{BCGPRKL2,TSchw08}.  If one uses the 3$\sigma$ ranges, one
  obtains $\sin^2\theta_{13}/\cos2\theta_{12}\ltap 0.22$ .}
$\cos2\theta_{12} \gg \sin^2\theta_{13}$.  In this case $|M_{ee}|$
does not depend on the Majorana phase $\alpha_{31}$ and on the Dirac
phase $\delta$.  From the measurement of the three observables,
$\BR_{ee}$, $\BR_{e\mu}$, and $\BR_{\mu\mu}$, three parameters, say,
$m_0$, $\alpha_{21}$ and $\alpha_{31}$, can, in principle, be
determined and information on the type of neutrino mass spectrum -
with NH, IH or QD can be 
obtained~\cite{Akeroyd:2007zv,Garayoa:2007fw,Kadastik:2007yd}.  This would
allow to tightly constrain $|M_{ee}|$.  Let us review briefly the
predictions for $\BR_{ee}$, $\BR_{e\mu}$, and $\BR_{\mu\mu}$ in the
cases of NH, IH and QD spectrum (see also
\cite{Akeroyd:2007zv,Garayoa:2007fw,Kadastik:2007yd}).

\subsection*{{\bf a) NH spectrum, ${\bf m_1 \ll m_2 < m_3}$.}}

Using the best fit values of the neutrino oscillation parameters one
finds \footnote{The limiting values quoted in this paragraph are
  obtained for the best fit values of the neutrinos oscillation
  parameters and for $\sin^22\theta_{13}\leq 0.14$.}  that in this
case $\BR_{ee}$, $\BR_{e\mu}$, and $\BR_{\mu\mu}$ can take values in
the following intervals: $0\ltap\BR^{\rm\sss NH}_{ee}\ltap 10^{-2}$, 
$0\ltap\BR^{\rm\sss NH}_{e\mu}\ltap0.08$, $0.16\ltap\BR^{\rm\sss NH}_{\mu\mu}\ltap0.31$.  
We get $\BR^{\rm\sss NH}_{ee} = 0$ for $(\alpha_{32} - 2\delta) = \pi$
and $s^2_{13}$ $= s^2_{12} \sqrt{\Delta m^2_{21}/\Delta m^2_{31}}$$\cong0.05$, 
while $\BR^{\rm\sss NH}_{e\mu} = 0$ for $(\alpha_{32} - \delta) = \pi$ and 
$s^2_{13}$$= s^2_{12}c^2_{12}\cot^2\theta_{23}(\Delta m^2_{21}/\Delta m^2_{31})$$\cong 6.9\times 10^{-3}$.  
The minimal and
maximal values of $\BR^{\rm\sss NH}_{\mu\mu}$ depend weakly on $s^2_{13}$.
Neglecting this dependence, one obtains a simple expression for the
Majorana phase (difference) $\alpha_{32}$ in terms of
$\BR^{\rm\sss NH}_{\mu\mu}$:
%%%%%%%%%%%%%%%%%%%%%%%%%%%%%%%%%%%%%
\begin{eqnarray}
 \cos \alpha_{32} \cong \frac{1}{2}\,
\left(\frac{\Delta m^2_{31}}{\Delta m^2_{21}}\right)^{\frac{1}{2}}
\frac{\BR^{\rm\sss NH}_{\mu\mu} - s^4_{23}}{c^2_{12}c^2_{23}s^2_{23}}
%\label{eq:Maj32}
\end{eqnarray}
%%%%%%%%%%%%%%%%%%%%%%%%%%%%%%%%%%%%
%
\subsection*{{\bf b) IH spectrum, ${\bf m_3 \ll m_1 < m_2}$.}}

We find very different results in this case:
$0.5c^4_{13}\cos^22\theta_{12} \ltap \BR^{\rm\sss IH}_{ee} \ltap 0.5c^4_{13}$
or $0.06 \ltap \BR^{\rm\sss IH}_{ee} \ltap 0.5$, 
$0\ltap \BR^{\rm\sss IH}_{e\mu}\ltap 0.48$, 
$0 \ltap \BR^{\rm\sss IH}_{\mu\mu}\ltap 0.14$.  One has
$\BR^{\rm\sss IH}_{\mu\mu} = 0$ for $\delta = 0$, $\alpha_{21} = \pi$ and
$s^2_{13}\cong 0.036$.  Now $\BR^{\rm\sss IH}_{ee}$ exhibits a very weak
dependence on $s^2_{13}$.  For the Majorana phase $\alpha_{21}$ we
obtain in terms of $\BR^{\rm\sss IH}_{ee}$:
%%%%%%%%%%%%%%%%%%%%%%%%%%%%%%%%%%%%%
\begin{eqnarray}
 \cos \alpha_{21} \cong 1 -
\frac{c^4_{13} - 2\BR^{\rm\sss IH}_{ee} }{2c^4_{13}c^2_{21}s^2_{21}}\,.
%\label{Maj32}
\end{eqnarray}
%%%%%%%%%%%%%%%%%%%%%%%%%%%%%%%%%%%%
%
\subsection*{{\bf c) QD spectrum, ${\bf m_{1,2,3} \gtap 0.1}$ eV.}}

It is not difficult to convince oneself that the branching ratios of
interest for the QD spectrum to a good approximation can take values
in the following intervals: $\cos^22\theta_{12}/3\ltap\BR^{\rm\sss QD}_{ee}\ltap 1/3$ 
or $0.03 \ltap \BR^{\rm\sss QD}_{ee} \ltap 0.33$, 
$0 \ltap\BR^{\rm\sss QD}_{e\mu}\ltap 0.46$, 
$\cos^22\theta_{23}/3 \ltap \BR^{\rm\sss QD}_{\mu\mu}\ltap 0.33$.  Actually, we 
have up to small
corrections $\BR^{\rm\sss QD}_{ee}\cong (2/3)\BR^{\rm\sss IH}_{ee}$.  For the Majorana
phase $\alpha_{21}$ in this case we get:
%%%%%%%%%%%%%%%%%%%%%%%%%%%%%%%%%%%%%
\begin{eqnarray}
 \cos \alpha_{21} \cong 1 -
\frac{c^4_{13} - 3\BR^{\rm\sss QD}_{ee} }{2c^4_{13}c^2_{21}s^2_{21}}\,.
%\label{Maj32}
\end{eqnarray}
%%%%%%%%%%%%%%%%%%%%%%%%%%%%%%%%%%%%
%
Given $\alpha_{21}$ and a sufficiently large $s_{13}$, information
about the Dirac phase $\delta$ and the Majorana phase $\alpha_{31}$
can be obtained from the knowledge of $\BR^{\rm\sss QD}_{e\mu}$ and
$\BR^{\rm\sss QD}_{\mu\mu}$.  If, however, a stringent limit on $s_{13}$ will
be obtained, $\alpha_{31}$ can be determined using $\BR^{\rm\sss QD}_{\mu\mu}$
and the knowledge of $\alpha_{21}$.

It is clear from the above simple analysis that the measurement of
the three branching ratios $\BR_{ee}$, $\BR_{e\mu}$, and
$\BR_{\mu\mu}$ would provide information about the type of neutrino
mass spectrum and the Majorana phases. If, for instance, it is
experimentally established that $\BR_{ee}> 0.01$, the neutrino mass
spectrum of NH type will be excluded.  The spectrum can either be of
IH or QD type.  If in addition $\BR_{\mu\mu}$ is determined to be
$\BR_{\mu\mu} > 0.14$, the IH spectrum will be ruled out.  If,
however, the neutrino mass spectrum will turn out to be QD, it will be
very difficult (if not practically impossible) to distinguish between
the spectrum with NO and that with IO, i.e. to get information about
the sign of $\Delta m^2_{31}$.

Consider next the more general case of $m_0$ having an arbitrary
value.  First, let us set $\theta_{13}=0$ for simplicity.  We will
consider the case of $\theta_{13}\neq 0$ later.  For $\theta_{13}=0$,
the main uncertainty in the prediction of $|M_{ee}|$ comes from the
lack of knowledge of $m_0$ and $\alpha_{21}$.  Note that in this case
$\BR_{ee}$ and $\BR_{e\mu}$ are independent of $\alpha_{31}$,
similarly to $M_{ee}$ ~\footnote{$\BR_{e\tau}$ is independent of
  $\alpha_{31}$ as well, but this mode is more difficult to measure
  than the two modes we are discussing.  }.  Knowing these two
branching ratios allows to determine $m_0$ and $\cos\alpha_{21}$. In
the case of spectrum with NO we have for the lightest neutrino mass:
%%%%%%%%%%%%%%%%%%%%%%%%%%%%%%%%%%%%%%%%%%%%%%%%%%
\begin{eqnarray}
\label{m1^2}
m_1^2
&=&
 \frac{
       ( \Delta m^2_{21} + \Delta m^2_{31} )
       ( 2 c_{23}^2 \BR_{ee} + \BR_{e\mu} )
       - 2 s_{12}^2 c_{23}^2 \Delta m^2_{21}
      }{
        2 c_{23}^2 - 6 c_{23}^2 \BR_{ee} - 3 \BR_{e\mu}
       }\,,\\
\cos\alpha_{21}
&=&
\frac{m_1}{\sqrt{m^2_1 + \Delta m^2_{21}}} +
 \frac{
       \Delta m^2_{21}( 2c^2_{12}c^2_{23} \BR_{ee} - s^2_{12}\BR_{e\mu})
       - m^2_1\BR_{e\mu}
      }{
         2 c_{12}^2 s_{12}^2 m_1 \sqrt{ m_1^2 + \Delta m^2_{21} }
        (\BR_{e\mu} + 2\BR_{ee}c^2_{23})
        }\,.
%\label{eq:mee-1}
\end{eqnarray}
%%%%%%%%%%%%%%%%%%%%%%%%%%%%%%%%%%%%%%%%%%
%
Equation (\ref{m1^2}) is valid also for the second to lightest
neutrino mass $m_1$ in the case of spectrum with IO.  The expression
for $\cos\alpha_{21}$ obviously cannot be used to determine
$\cos\alpha_{21}$ for $m_1 = 0$: for $\theta_{13} = 0$ and $m_1=0$,
neither $\BR_{ee}$ nor $\BR_{e\mu}$ depend on $\alpha_{21}$.  In the
case of spectrum with IO (inverted ordering) we obtain:
%%%%%%%%%%%%%%%%%%%%%%%%%%%%%%%%%%%%%%%%%%%%%%%%%%
\begin{eqnarray}
\label{m3^2IO}
m_3^2
&=&
 \frac{
       (2\Delta m^2_{23} - \Delta m^2_{21} )
       ( 2 c_{23}^2 \BR_{ee} + \BR_{e\mu} )
       - 2 c_{23}^2(\Delta m^2_{23 }- c^2_{12}\Delta m^2_{21})
      }{
        2 c_{23}^2 - 6 c_{23}^2 \BR_{ee} - 3 \BR_{e\mu}
       }\,,\\
\label{al21IO}
\cos\alpha_{21}
&=& 1 - \frac{\BR_{e\mu}}{2c_{12}^2 s_{12}^2(\BR_{e\mu} + 2\BR_{ee}c^2_{23})} +
             O\left(\frac{\Delta m^2_{21}}{\Delta m^2_{23}}\right)\,.
\end{eqnarray}
%%%%%%%%%%%%%%%%%%%%%%%%%%%%%%%%%%%%%%%%%%
%
  As can be expected, for $m^2_1 \gg \Delta m^2_{21}$, the expression
for $\cos\alpha_{21}$ in the case of spectrum with NO coincides with
that for spectrum with IO. Using eq.~(\ref{m1^2}), we get a universal
expression for $|M_{ee}|$ valid for both types of spectrum - with NO
and IO and any hierarchy between neutrino masses:
%%%%%%%%%%%%%%%%%%%%%%%%%%%
\begin{eqnarray}
\label{eq:mee-0}
|M_{ee}|^2
=
\left( \sum_i m_i^2 \right) \BR_{ee}
&=&
 \frac{
       2 c_{23}^2 \Delta m^2_{31}
       - 2 c_{23}^2 ( 3 s_{12}^2 - 1 ) \Delta m^2_{21}
      }{
        2 c_{23}^2 - 6 c_{23}^2 \BR_{ee} - 3 \BR_{e\mu}
       } \BR_{ee}\\
&\simeq&
 \frac{
       \text{sgn}(\Delta m^2_{31}) \times \BR_{ee}
      }{
        1 - 3 \BR_{ee} - 3 \BR_{e\mu}
       }
     \times 2.4\times 10^{-3} \eV^2,
\label{eq:mee-1}
\end{eqnarray}
%%%%%%%%%%%%%%%%%%%%%%%%%%%%%%%
%
where in the last equation we have used the best fit value of
$\theta_{23}$ and have neglected the term $\sim (3s^2_{12} - 1)\Delta m^2_{21}/\Delta m^2_{31}$.  
Note that the denominator in the
expression for $|M_{ee}|$, eq.~(\ref{eq:mee-0}), does not go through
zero since we have:
%%%%%%%%%%%%%%%%%%%%%%%%%%%%%%%%%%%%%%%%%%%%%%%%%%
\begin{eqnarray}
\label{ee+emuNO}
2 c_{23}^2 \BR_{ee} + \BR_{e\mu} =
 \frac{2 c_{23}^2
      (m^2_{1} + s^2_{12}\Delta m^2_{21})}
       {3m^2_1 + \Delta m^2_{21} + \Delta m^2_{31}}
       \,,~~{\rm NO~spectrum}\,,\\
2 c_{23}^2 \BR_{ee} + \BR_{e\mu} =
 \frac{2 c_{23}^2
      (m^2_{3} + \Delta m^2_{23})}
       {3m^2_3 + 2\Delta m^2_{23}}
       \,,~~{\rm IO~spectrum}\,,
\label{ee+emuIO}
\end{eqnarray}
%%%%%%%%%%%%%%%%%%%%%%%%%%%%%%%%%%%%%
%
where we have neglected terms $\sim (\Delta m^2_{21}/\Delta m^2_{23})$
in the second equation. We see that in the QD region, where
$|M_{ee}|$ has a relatively large value, one has
$(2c^2_{23}\BR_{ee}+\BR_{e\mu})\cong(2/3)c^2_{23}(1 + \Delta m^2_{23}/(3m^2_0))$.

It follows from eq.~(\ref{eq:mee-1}) that $|M_{ee}| > 0.05~\eV \simeq
\sqrt{|\Delta m^2_{31}|}$ can be predicted without the knowledge of
$\text{sgn}(\Delta m^2_{31})$, if the collider experiments show that
the branching ratios $\BR_{ee}$ and $\BR_{e\mu}$ satisfy
%%%%%%%%%%%%%%%%%%%%%%%%%%%%%%
\begin{eqnarray}
 -\frac{4 c_{23}^2}{\,3\,} \BR_{ee} + \frac{2 c_{23}^2}{\,3\,}
 \gtrsim \BR_{e\mu}
 \gtrsim -\frac{8 c_{23}^2}{\,3\,} \BR_{ee} + \frac{2 c_{23}^2}{\,3\,}
\end{eqnarray}
%%%%%%%%%%%%%%%%%%%%%%%
%
If indeed $\sin^2\theta_{13}$ is negligibly small and these conditions
are satisfied by the measured $\BR_{ee}$ and $\BR_{e\mu}$, a
positive result can be expected in the next generation of
$\betabeta$-decay experiments having a sensitivity to $|M_{ee}| \geq0.05$~eV.  Note 
that the magnitude of the left and right sides of the
inequality is very sensitive to the value of $c^2_{23}$. Note also
that these conditions do not depend explicitly on $\BR_{\mu\mu}$. For
this reason we will first obtain constraints on $|M_{ee}|$ using only
the branching ratios $\BR_{ee}$ and $\BR_{e\mu}$.

Next, we analyse the case of $\theta_{13}\neq 0$ numerically.  We
calculated $\BR_{ee}$, $\BR_{e\mu}$, and $M_{ee}$ by using $|\Delta m^2_{31}|$, 
$\Delta m^2_{21}$, and $\sin^2{2\theta_{12}}$ given in
eq.~(\ref{eq:solvalues}) and (\ref{eq:atmvalues}).  We allow $m_0$ to
vary in the interval in eq.~(\ref{minmj}), while the other parameters
are varied in the following ranges reflecting the uncertainties in
their knowledge or lack of any constraints:
%%%%%%%%%%%%%%%%%%%%%%%%%%%%%%%%%%%%%
\begin{eqnarray}
\sin^2{2\theta_{23}} > 0.94, \ \
\sin^2{2\theta_{13}} < 0.14, \ \
\delta, \alpha_{21}, \alpha_{31} = 0\text{-}2\pi.
\label{ranges}
\end{eqnarray}
%%%%%%%%%%%%%%%%%%%%%%%%%%%%%%%%%%
%
Later we will present results for the prospective smaller
uncertainties in $\sin^2{2\theta_{23}}$ and $\sin^2{2\theta_{13}}$,
corresponding to $\sin^2{2\theta_{23}} > 0.99$ and
$\sin^2{2\theta_{13}} < 0.04$.

\begin{figure}
\begin{center}
\includegraphics[angle=-90,width=5cm]{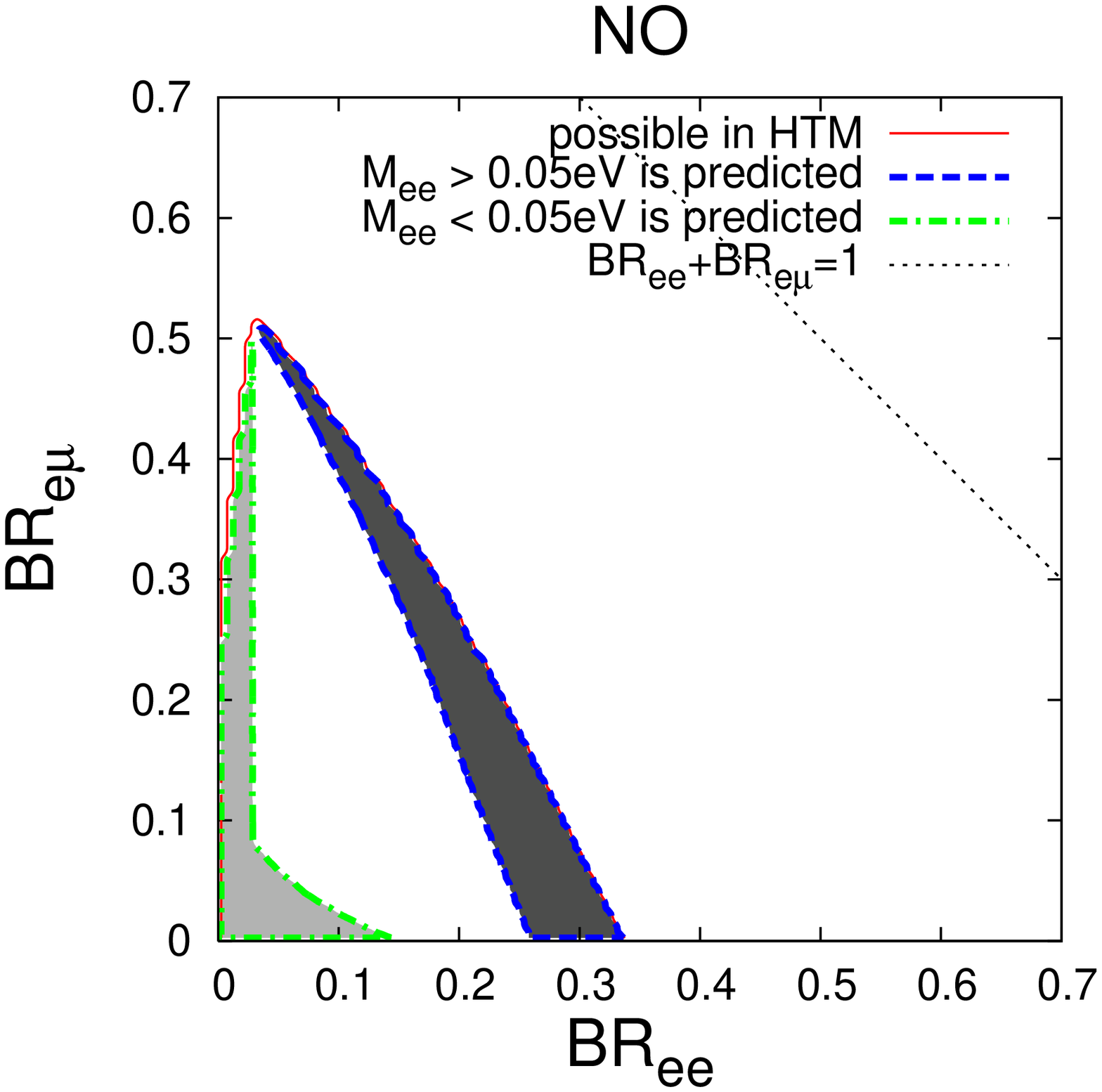}
\includegraphics[angle=-90,width=5cm]{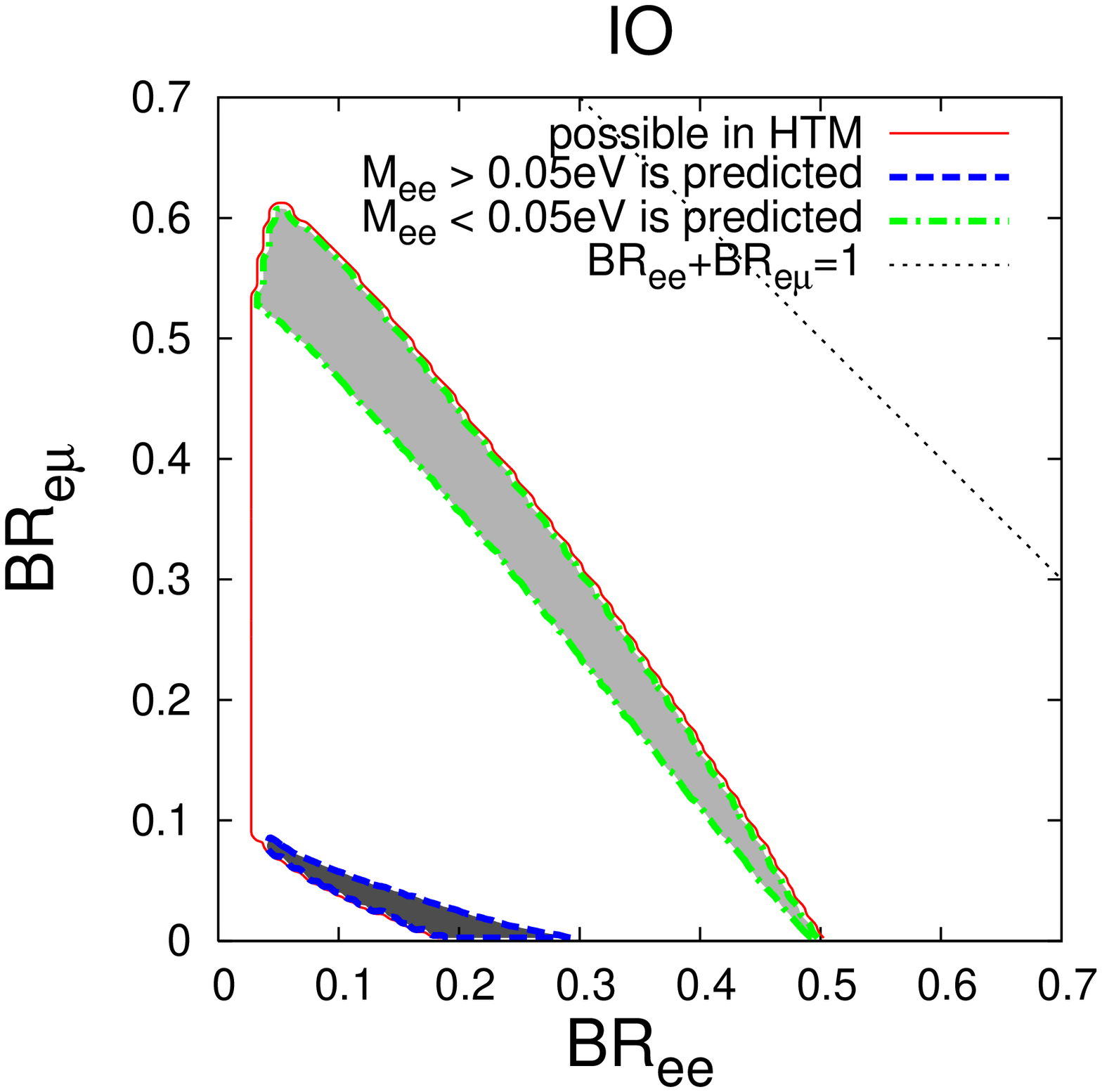}
\includegraphics[angle=-90,width=5cm]{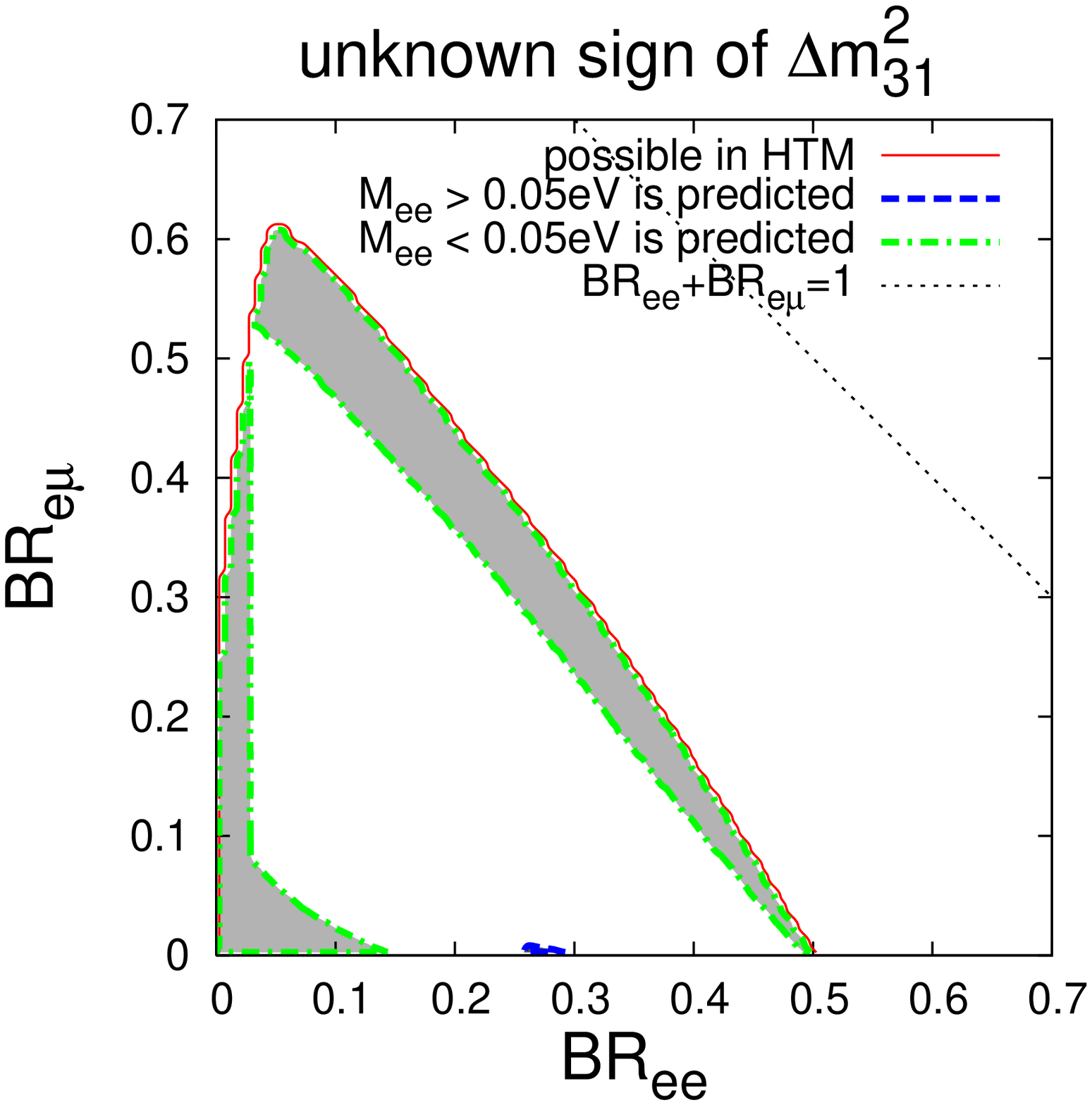}
\caption{Values of $\BR_{ee}$ and $\BR_{e\mu}$, for which $|M_{ee}| >
  0.05~\eV$ (black areas limited by the dashed blue lines) or
  $|M_{ee}| < 0.05~\eV$ (grey areas limited by the dash-dotted green
  lines) in the HTM. The solid (red) line shows the entire region of
  allowed values of $\BR_{ee}$ and $\BR_{e\mu}$ in the HTM (black and
  grey areas and the white area between the coloured one).  The
  results shown are obtained by varying $m_0$, $\sin^2{2\theta_{23}}$,
  $\sin^2{2\theta_{13}}$, $\delta$, $\alpha_{21}$ and $\alpha_{31}$,
  in the ranges given in eqs.~(\ref{minmj}) and (\ref{ranges}).  The
  left and middle panels correspond to NO and IO spectrum,
  respectively, while the right panel was obtained assuming that 
  ${\rm sgn}(\Delta m^2_{31})$ is unknown. The dotted line corresponds to
  $\BR_{ee}+\BR_{e\mu} = 1$.  The region above this line is
  unphysical.  See text for further details.}
\label{fig:noBRmm}
\end{center}
\end{figure}

In Fig.~\ref{fig:noBRmm} we show the regions in the $\BR_{ee}-\BR_{e\mu}$ 
plane where we {\it definitely} have $|M_{ee}| \geq0.05~\eV$ 
or $|M_{ee}| < 0.05~\eV$. More specifically, the solid (red)
line determines the complete allowed region in the HTM, corresponding
to $\Delta m^2_{21}$, $|\Delta m^2_{31}|$ and $\sin^22\theta_{12}$
given in eqs.~(\ref{eq:solvalues}) and (\ref{eq:atmvalues}), and
values of $m_0$, $\sin^2{2\theta_{23}}$, $\sin^2{2\theta_{13}}$,
$\delta$, $\alpha_{21}$ and $\alpha_{31}$, which were allowed to vary
in the ranges specified in eqs.~(\ref{minmj}) and (\ref{ranges}).  The
dashed blue (dash-dotted green) lines determine the black (grey)
regions where $|M_{ee}|$ is {\it definitely} larger (smaller) than
$0.05~\eV$ in the HTM when $m_0$, $\sin^2{2\theta_{23}}$,
$\sin^2{2\theta_{13}}$, $\delta$, $\alpha_{21}$ and $\alpha_{31}$, are
varied within the indicated intervals (i.e. eqs.~(\ref{minmj}) and
(\ref{ranges})).  For values of $\BR_{ee}$ and $\BR_{e\mu}$ from the
region depicted in white (and located between those shown in black and
in grey), the determination of $|M_{ee}|$ is not unambiguous: both
values of $|M_{ee}|\geq0.05~\eV$ and $|M_{ee}| < 0.05~\eV$ are
possible.  This degeneracy can be lifted to certain extent, but not
completely, by using additional information on $\BR_{\mu\mu}$ (see
further).  The dotted black line in Fig.~\ref{fig:noBRmm} corresponds
to $\BR_{ee}+\BR_{e\mu} = 1$.  We show it only to indicate the
boundary of the region of possible values of $\BR_{ee}$ and
$\BR_{e\mu}$: the area above this line is unphysical.  The results in
Fig.~\ref{fig:noBRmm} are obtained without using the possible
additional data on \footnote{We recall that in this analysis we do not
  use possible data on $\BR_{\tau\mu}$, $\BR_{\tau\tau}$ and
  $\BR_{e\tau}$.}  $\BR_{\mu\mu}$.  The left and middle panels
correspond to NO and IO spectrum, respectively, while the results
shown in the right panel were obtained assuming that the 
${\rm sgn}(\Delta m^2_{31})$ (i.e. the type of the neutrino mass
spectrum) is unknown. The black area where $|M_{ee}|$ is, e.g. 
{\it definitely} larger than $0.05~\eV$ in the right panel corresponds to
the intersection of the black areas in left and middle panels. Note
that we can have $|M_{ee}| \gtap 0.05~\eV$ also in the region shown in
white and located between the grey areas in the right panel of
Fig.~\ref{fig:noBRmm}.  This cannot be unambiguously predicted,
however, knowing only the values of $\BR_{ee}$ and $\BR_{e\mu}$ which
lie in the white area.

\begin{figure}
\begin{center}
\includegraphics[angle=-90,width=5cm]{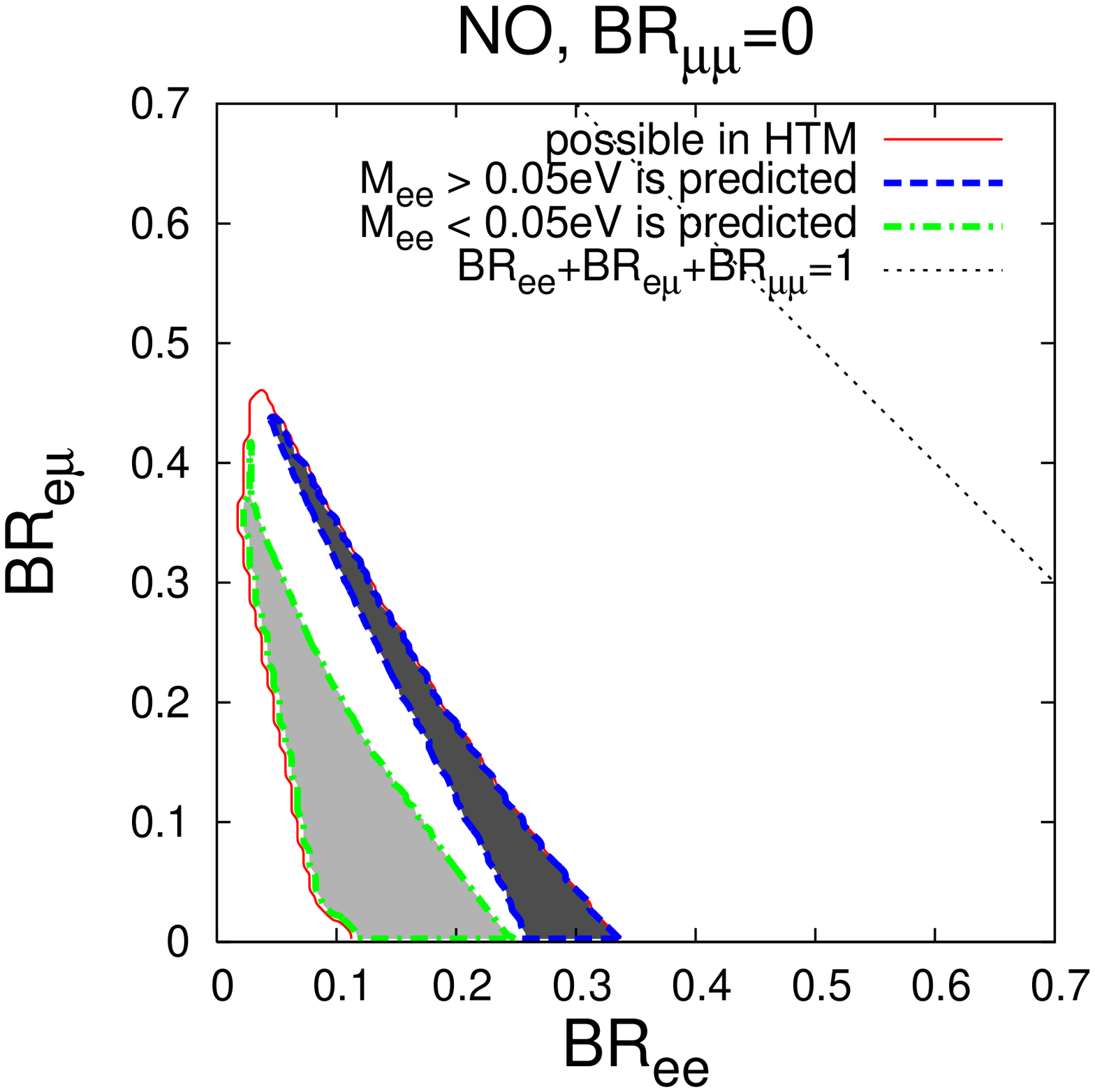}
\includegraphics[angle=-90,width=5cm]{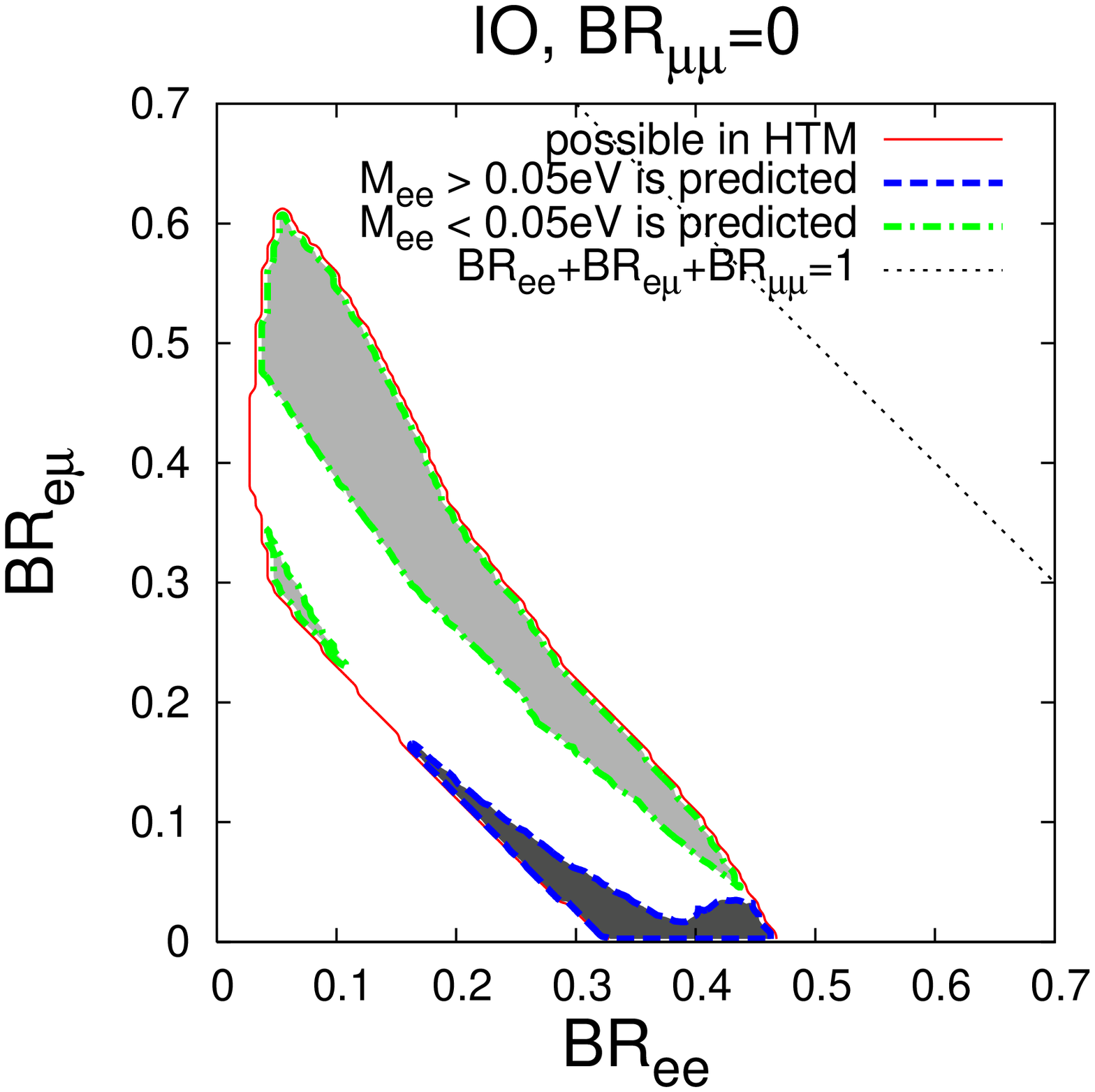}
\includegraphics[angle=-90,width=5cm]{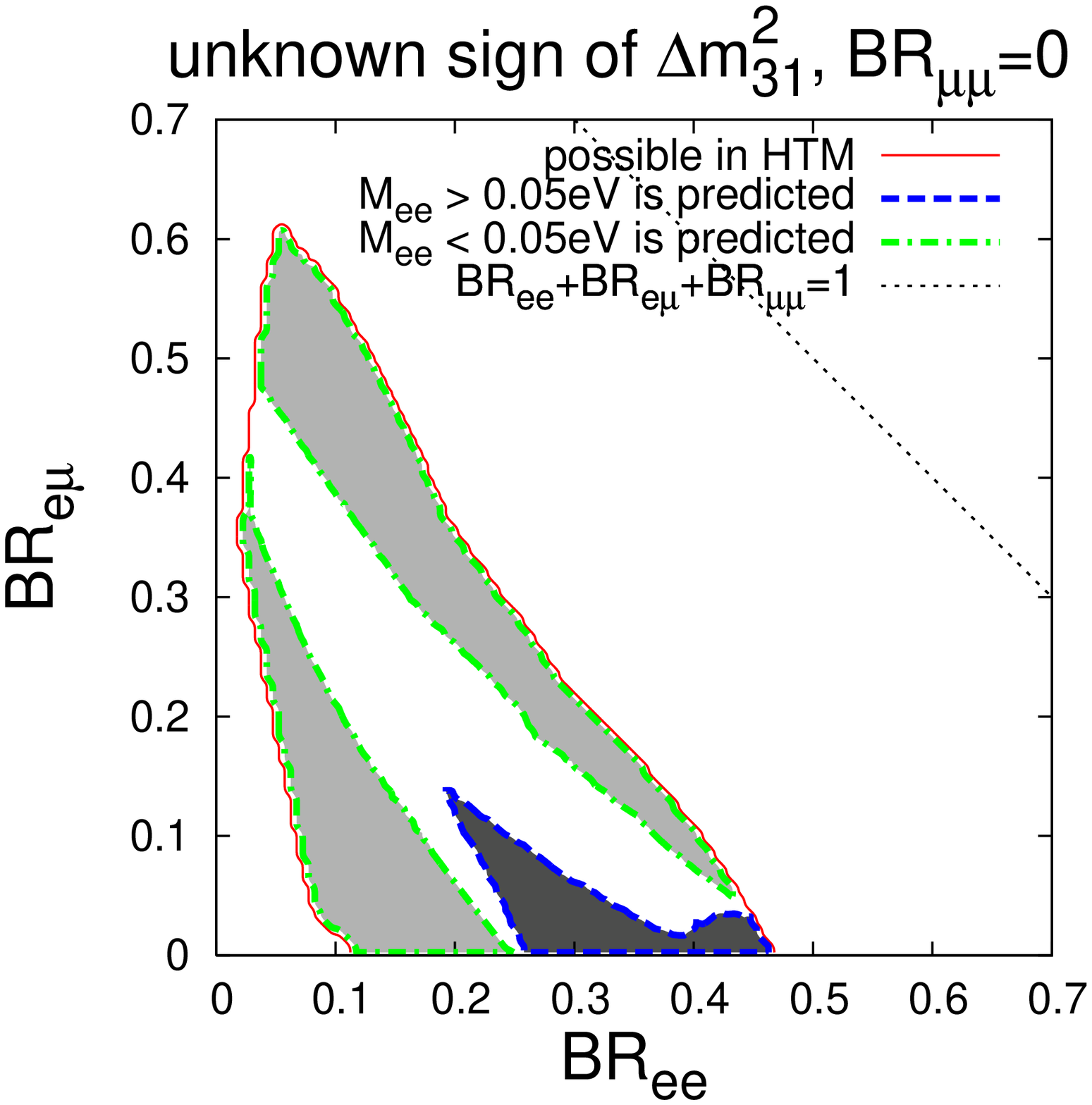}
\caption{The same as in Fig.~\ref{fig:noBRmm}, but assuming that the
  experimentally determined $\BR_{\mu\mu}=0$.  The dotted line
  corresponds to $\BR_{ee}+\BR_{e\mu}+\BR_{\mu\mu} = 1$; the region
  above the line is unphysical.  See text for further details.}
\label{fig:BRmm0}
\end{center}
\end{figure}

\begin{figure}
\begin{center}
\includegraphics[angle=-90,width=5cm]{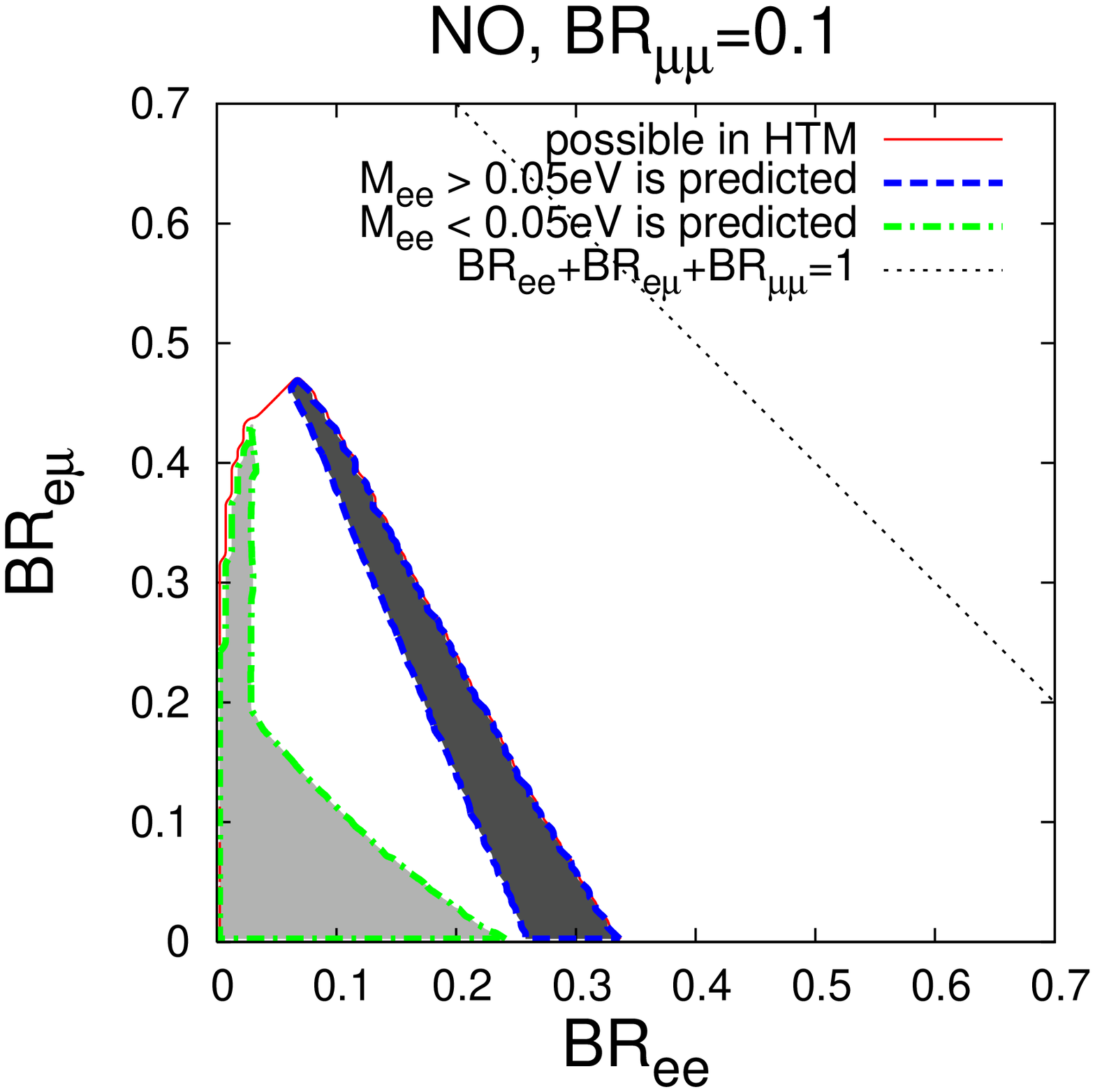}
\includegraphics[angle=-90,width=5cm]{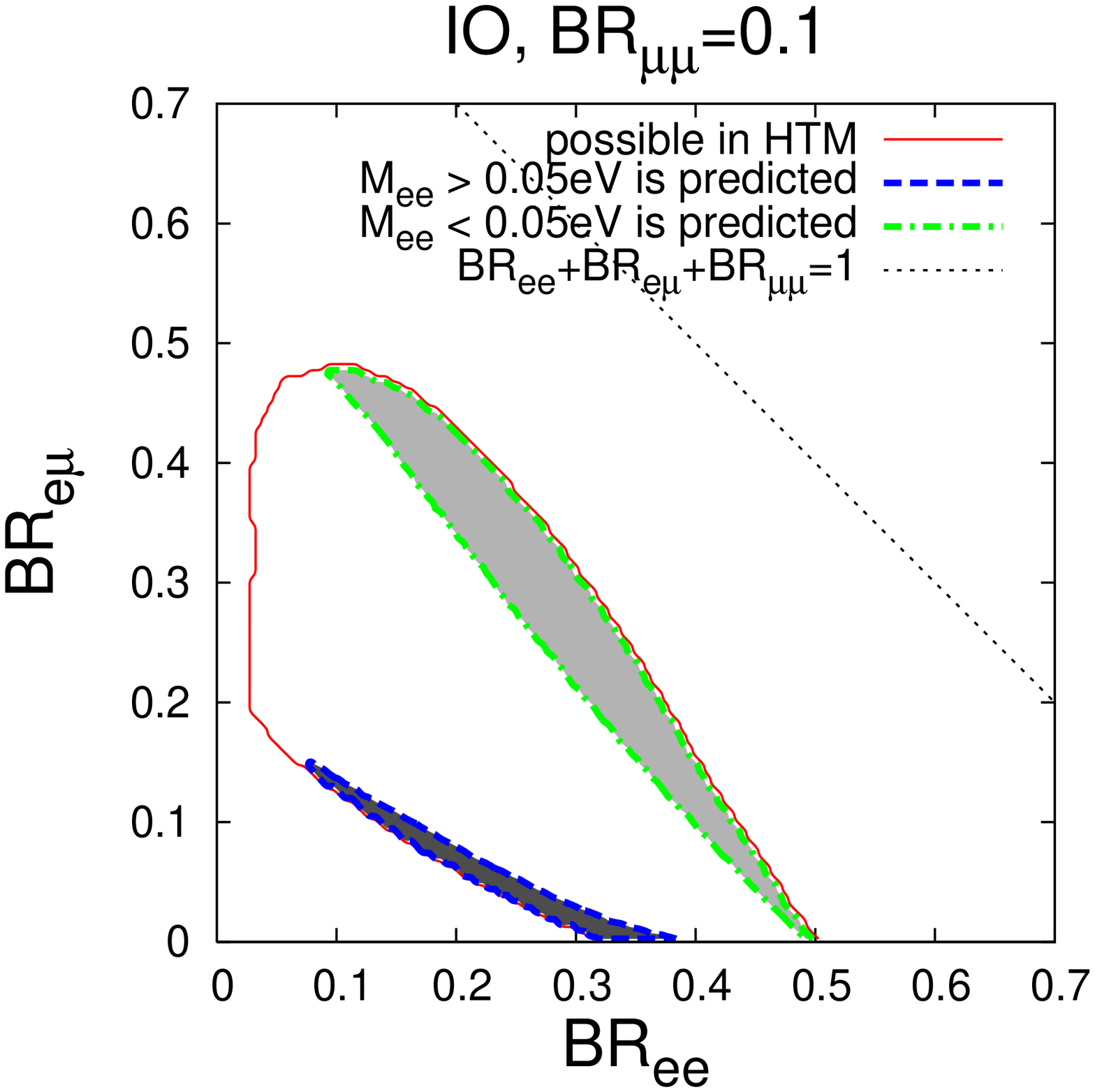}
\includegraphics[angle=-90,width=5cm]{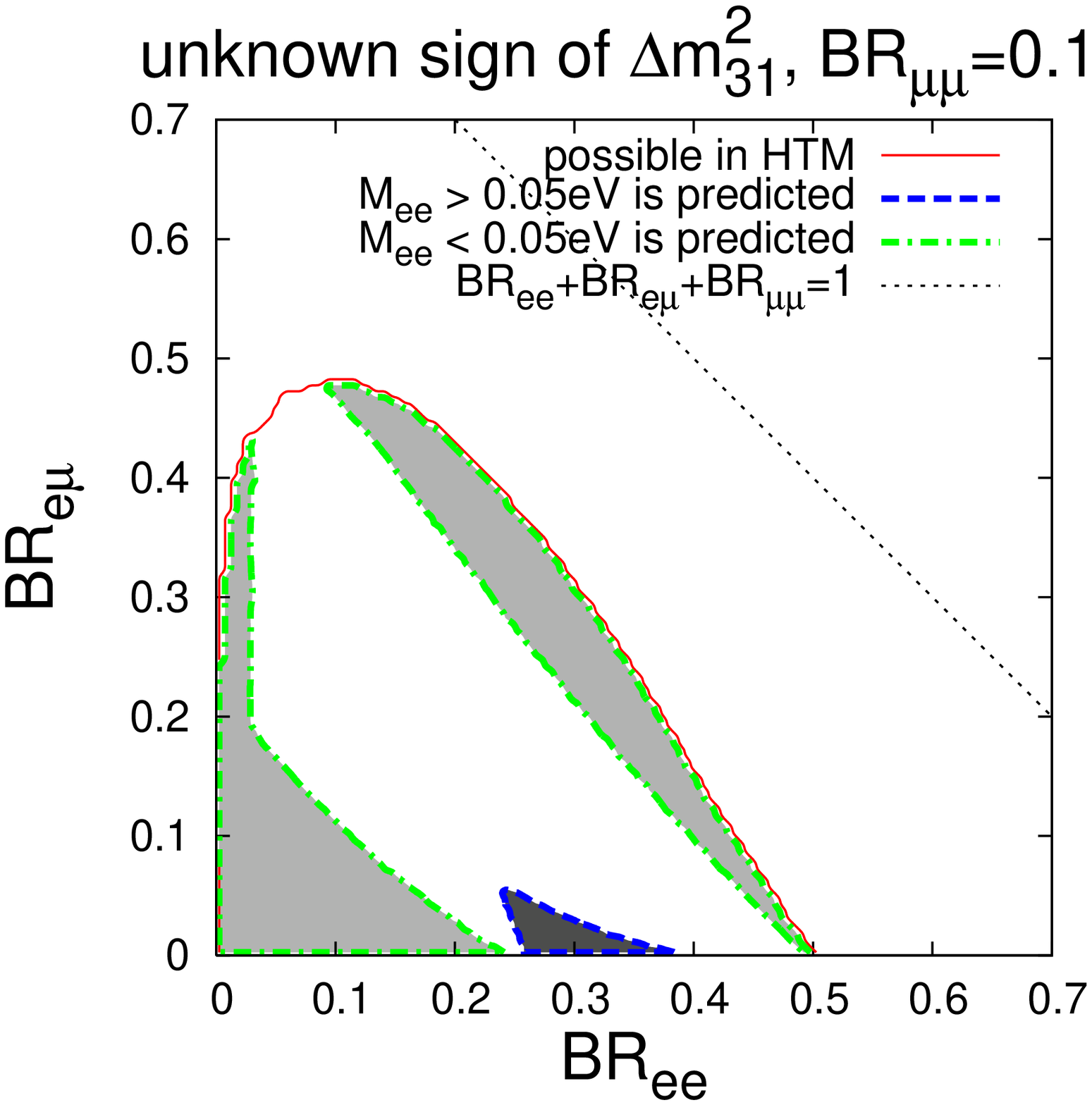}
\caption{The same as in Fig.~\ref{fig:BRmm0}, but for $\BR_{\mu\mu}=0.1$.}
\label{fig:BRmm01}
\end{center}
\end{figure}

\begin{figure}
\begin{center}
\includegraphics[angle=-90,width=5cm]{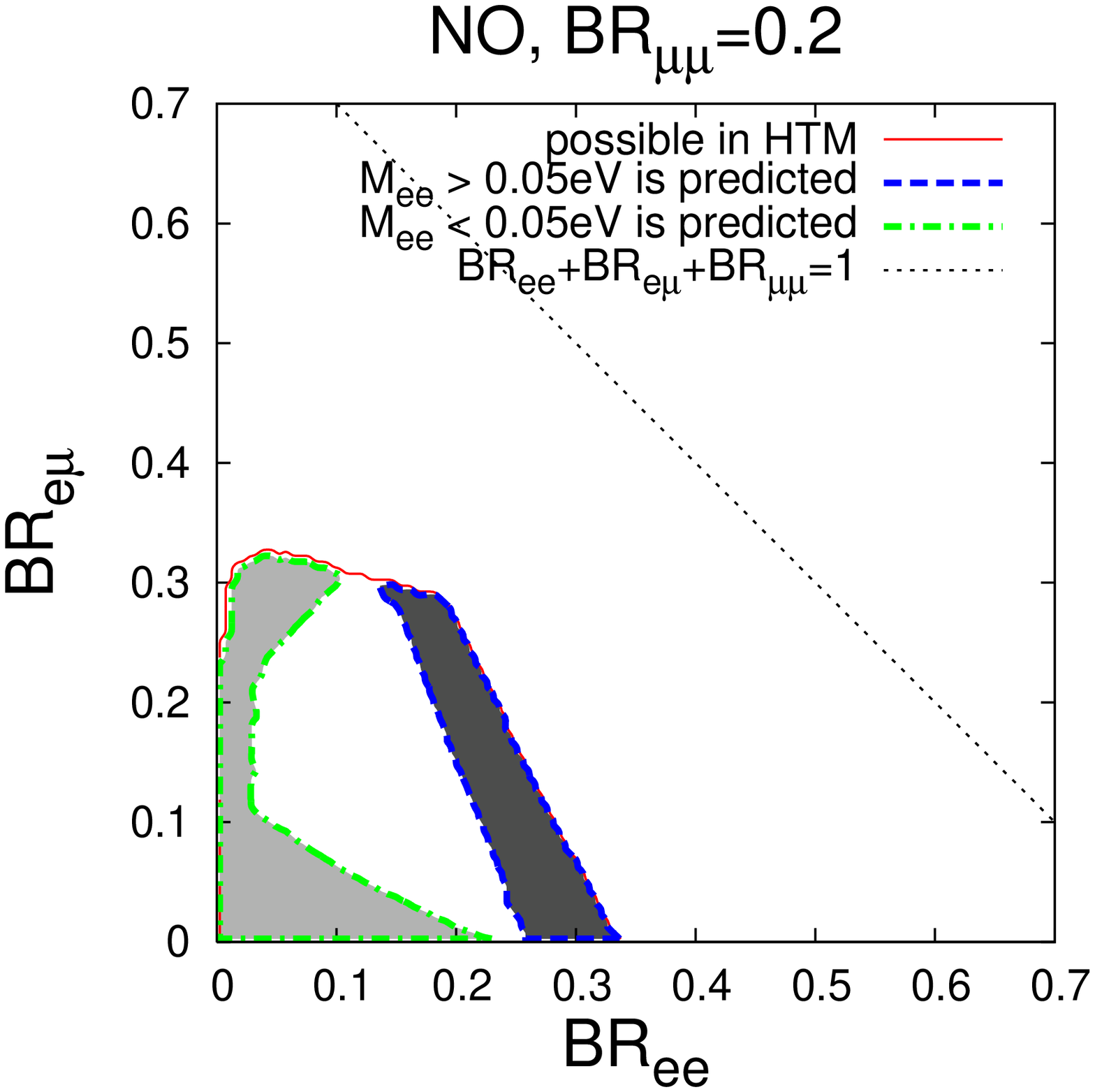}
\includegraphics[angle=-90,width=5cm]{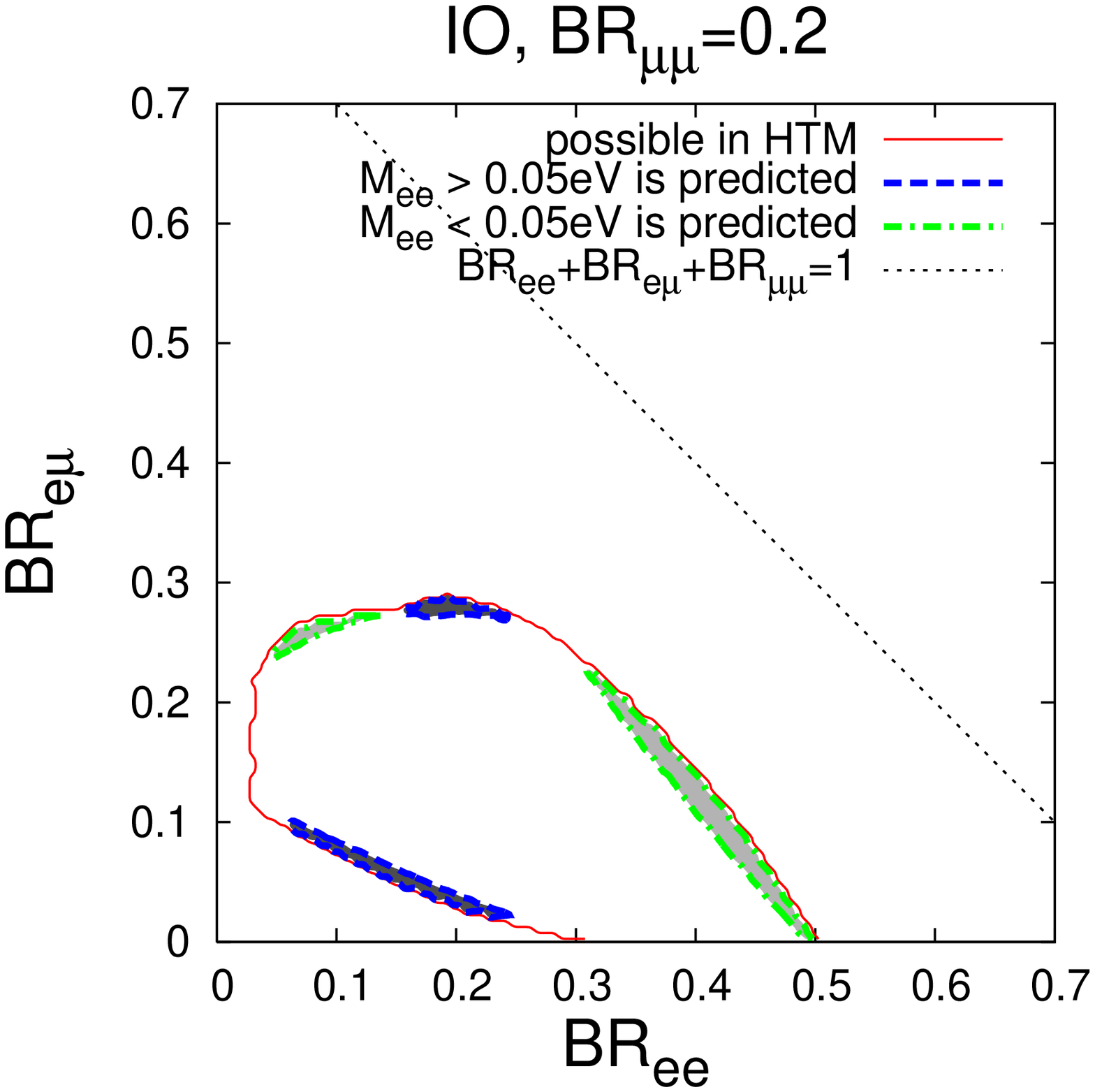}
\includegraphics[angle=-90,width=5cm]{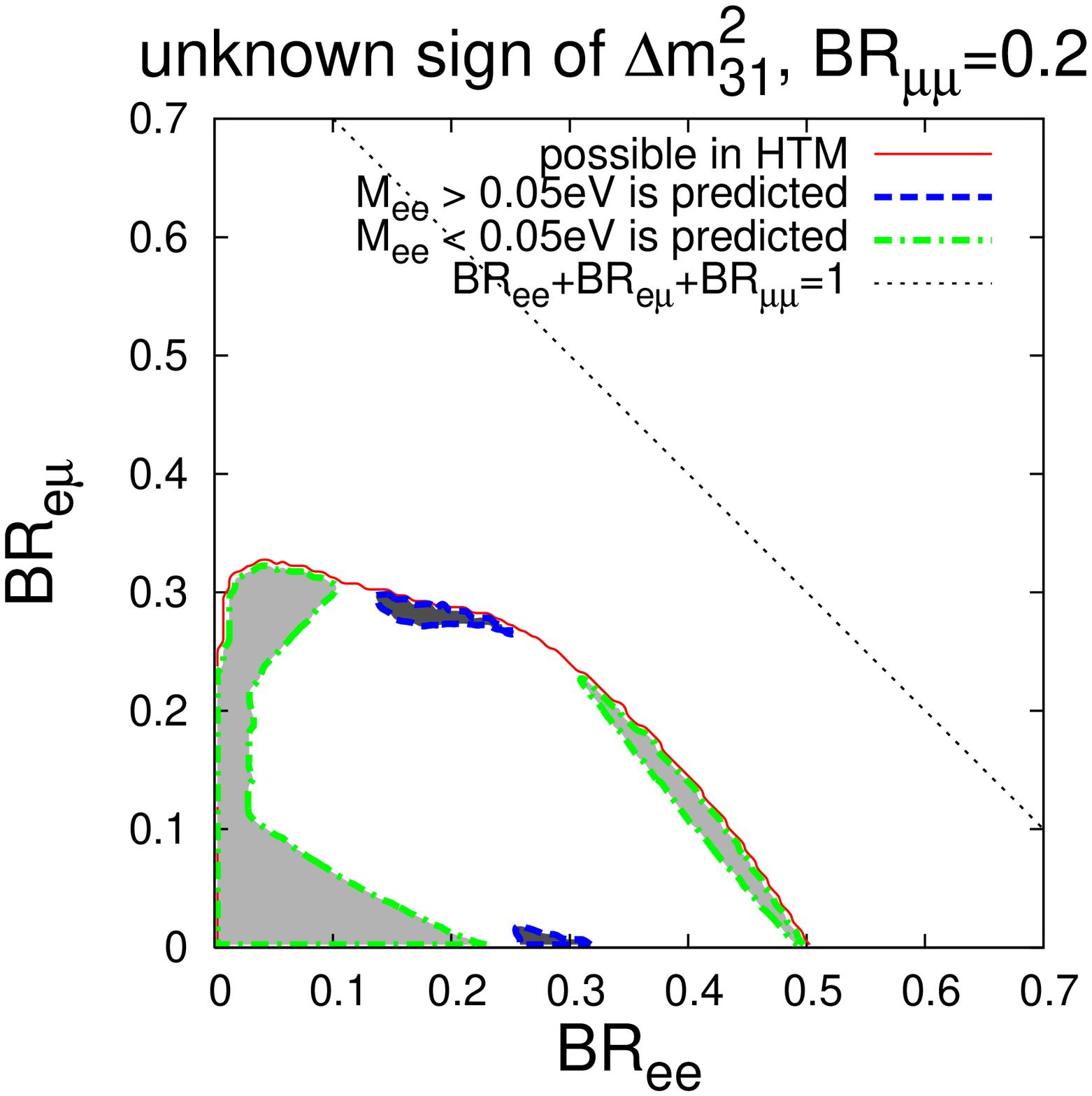}
\caption{The same as in Fig.~\ref{fig:BRmm0}, but for $\BR_{\mu\mu}=0.2$.}
\label{fig:BRmm02}
\end{center}
\end{figure}

\begin{figure}
\begin{center}
\includegraphics[angle=-90,width=5cm]{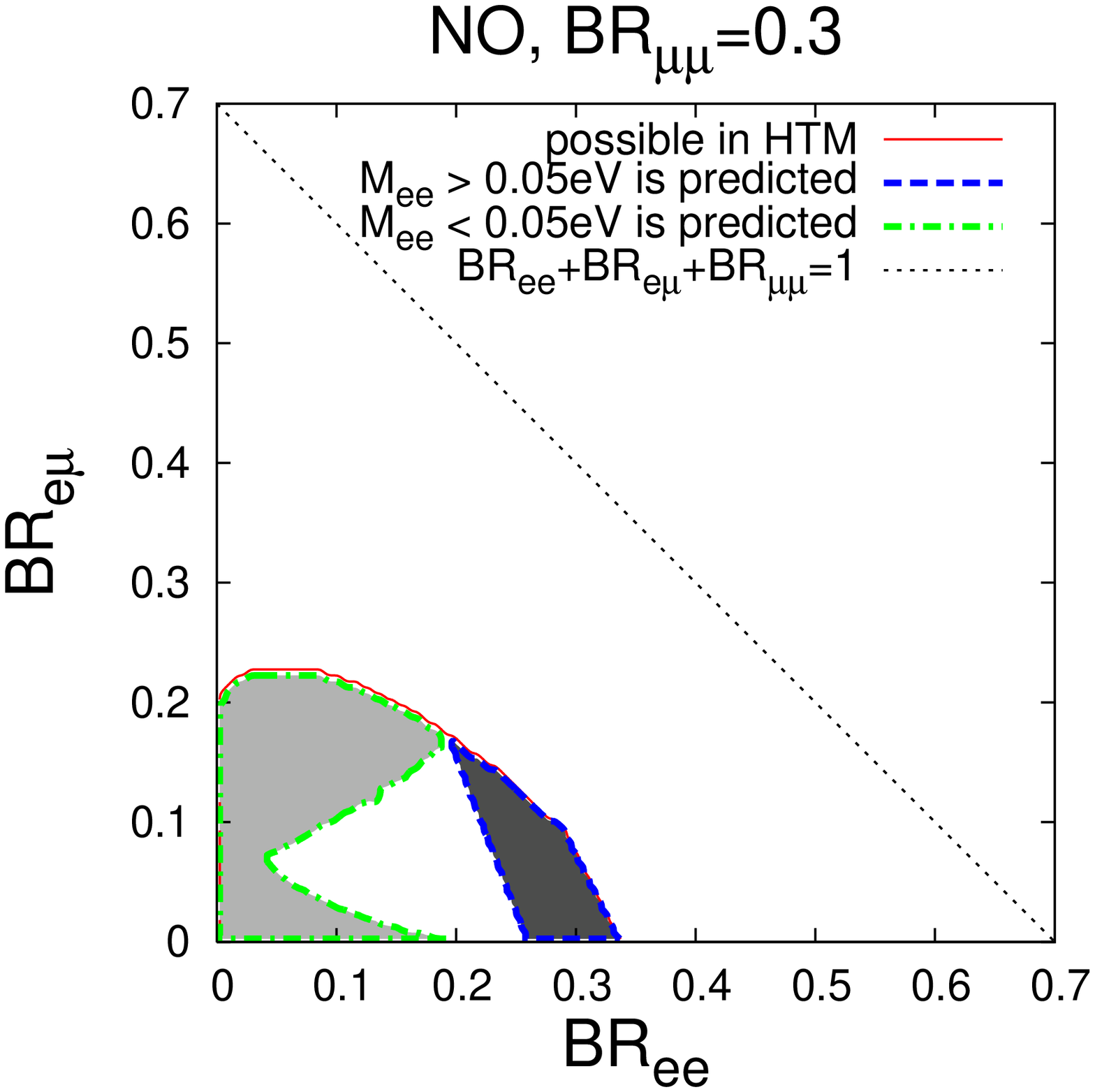}
\includegraphics[angle=-90,width=5cm]{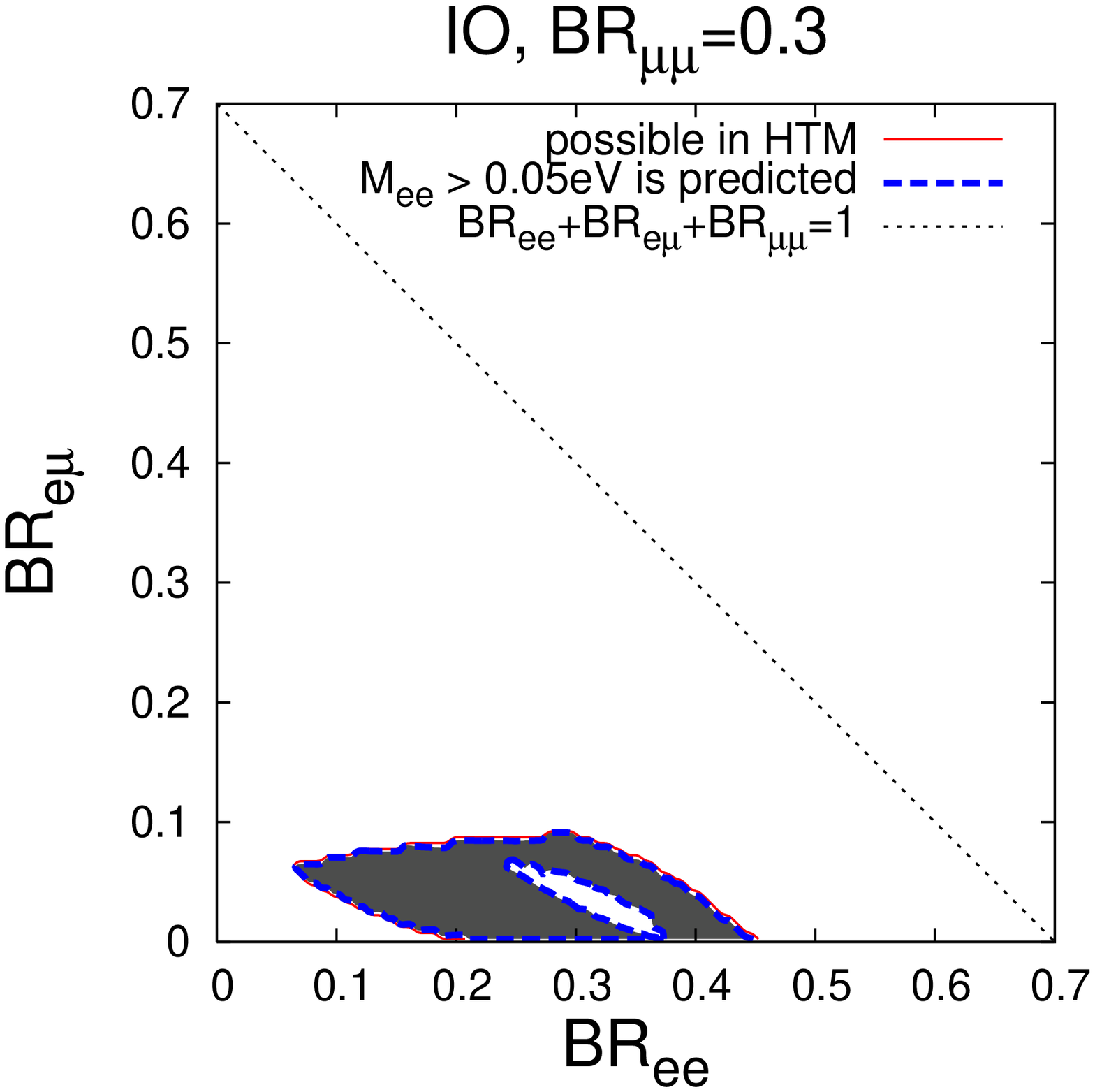}
\includegraphics[angle=-90,width=5cm]{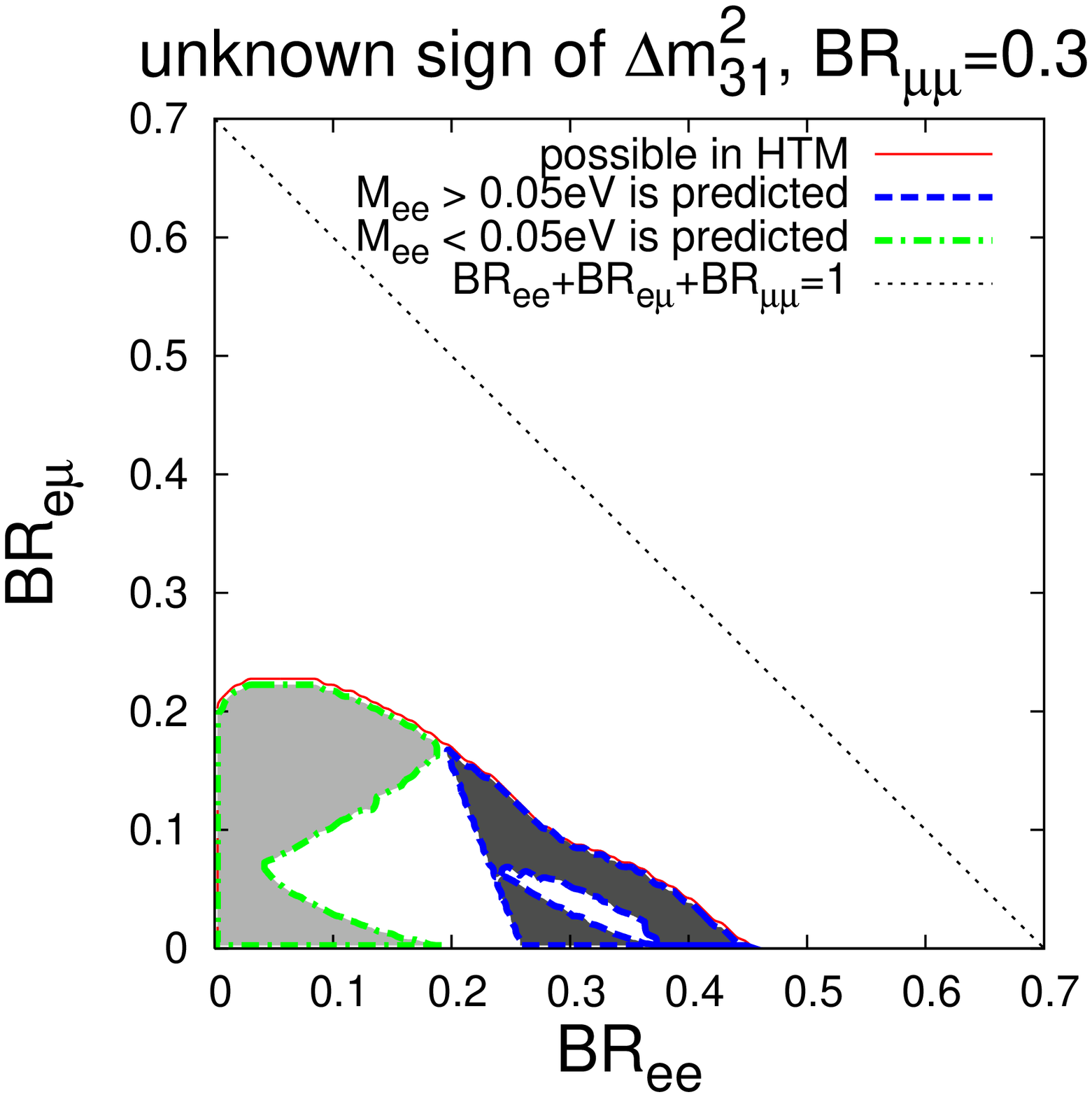}
\caption{The same as in fig.~\ref{fig:BRmm0}, but for $\BR_{\mu\mu}=0.3$.}
\label{fig:BRmm03}
\end{center}
\end{figure}

Next we show how the results discussed above change when we add
information on $\BR_{\mu\mu}$. The $\mu^{\pm}\mu^{\pm}$ decay mode of
$H^{\pm\pm}$ is relatively easy to measure at LHC by virtue of the two
same sign muons in the final state.  We present results for
$\BR_{\mu\mu}= 0;~0.1;~0.2;~0.3$ in Figs.~\ref{fig:BRmm0},
\ref{fig:BRmm01}, \ref{fig:BRmm02} and \ref{fig:BRmm03}, respectively,
where $\BR_{\mu\mu}= 0$ in practice corresponds to $\BR_{\mu\mu} <0.01$.  
When we quote a
%%%%%%%%%%%%%%%%%%%%%%%%%
%
%
%%%%%%%%%%%%%%%%%%%%%%%%%%
% \newpage
%%%%%%%%%%%%%%%%%%%%%%%%%
%%%%%%%%%%%%%%%%%%%%%%%%%%
\begin{figure}
\begin{center}
\includegraphics[angle=-90,width=5cm]{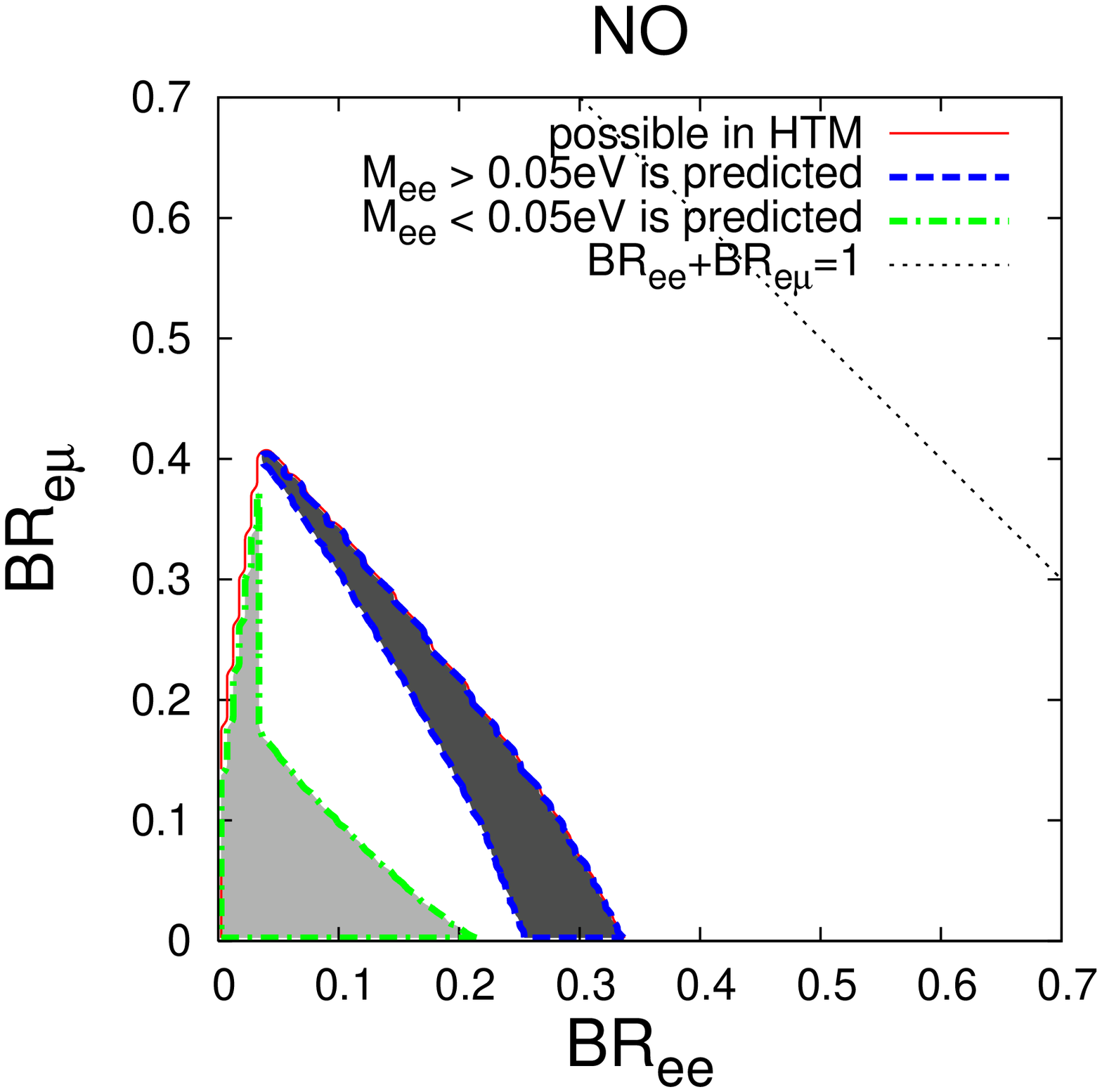}
\includegraphics[angle=-90,width=5cm]{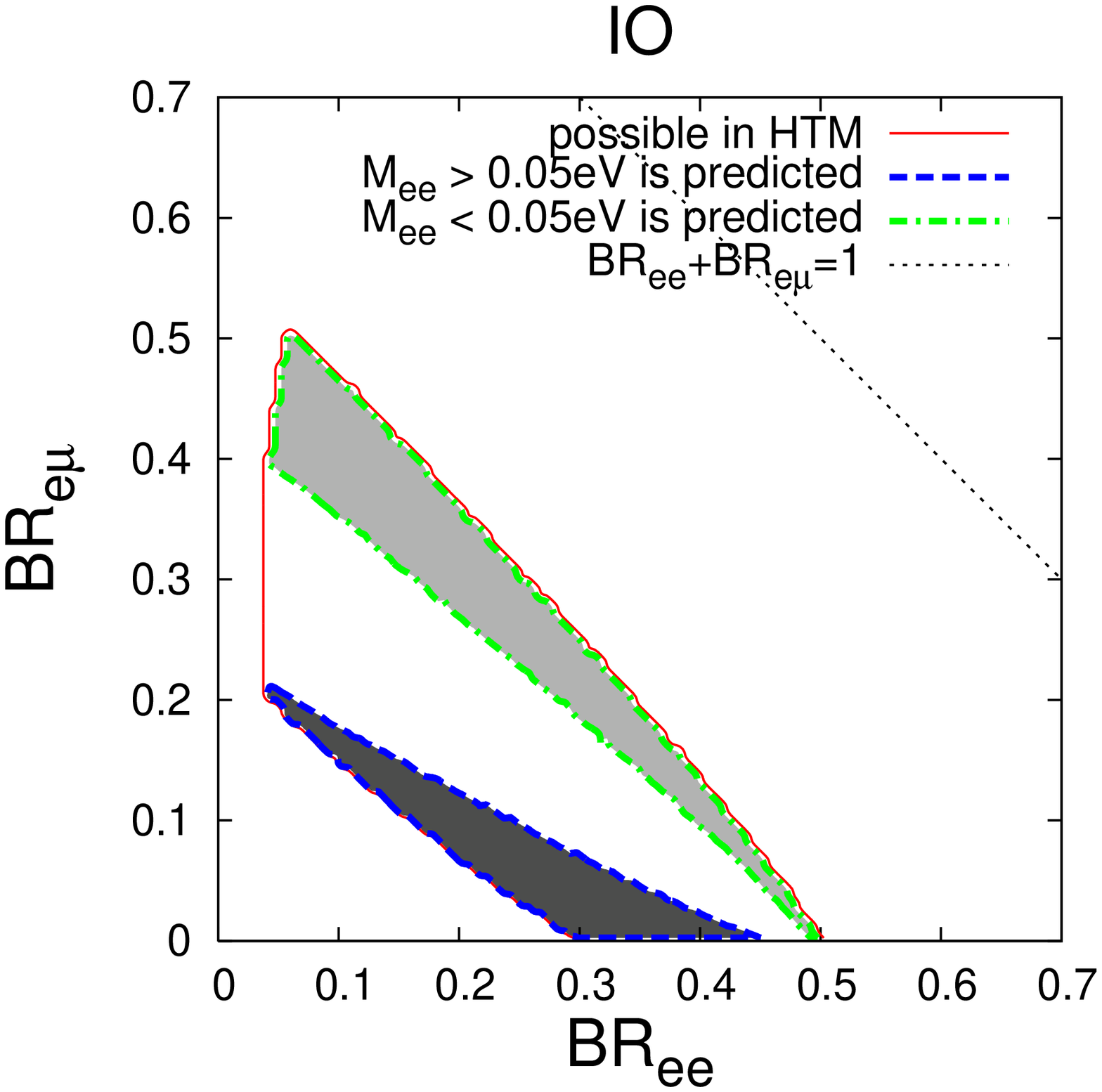}
\includegraphics[angle=-90,width=5cm]{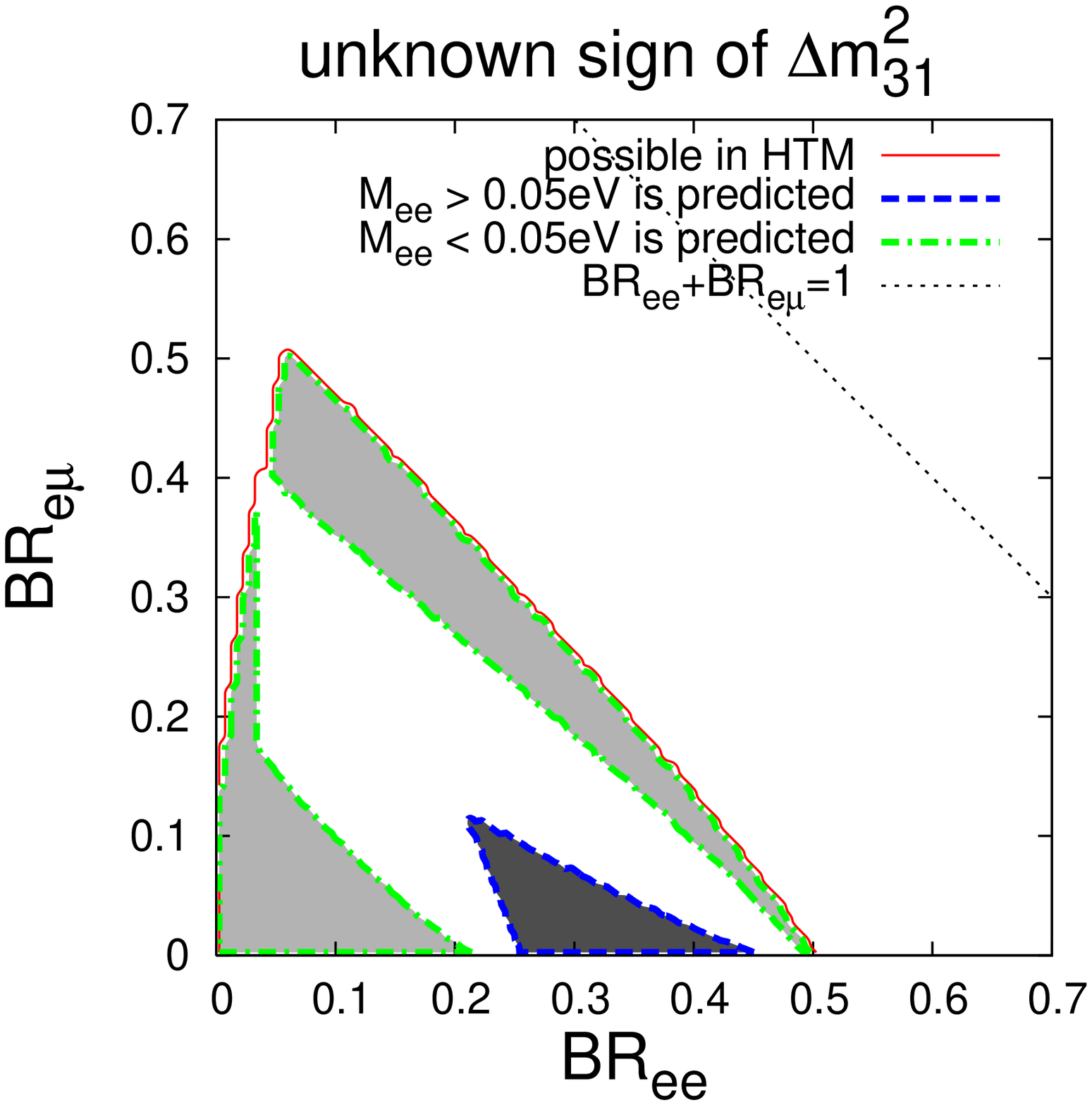}
\caption{The same as in Fig.~\ref{fig:noBRmm}, but allowing
  $\sin^2{2\theta_{23}}$ and $\sin^2{2\theta_{13}}$ to vary in the
  following more narrow intervals: $\sin^2{2\theta_{23}} > 0.99$ and
  $\sin^2{2\theta_{13}} < 0.04$.  No information on $\BR_{\mu\mu}$ was
  used in deriving the results shown in the figure.  See text for
  further details.}
\label{fig:noBRmmimpr}
\end{center}
\end{figure}
%%%%%%%%%%%%%%%%%%%%%%%%%%%%%%%%%%%%%%%%
\begin{figure}
\begin{center}
\includegraphics[angle=-90,width=5cm]{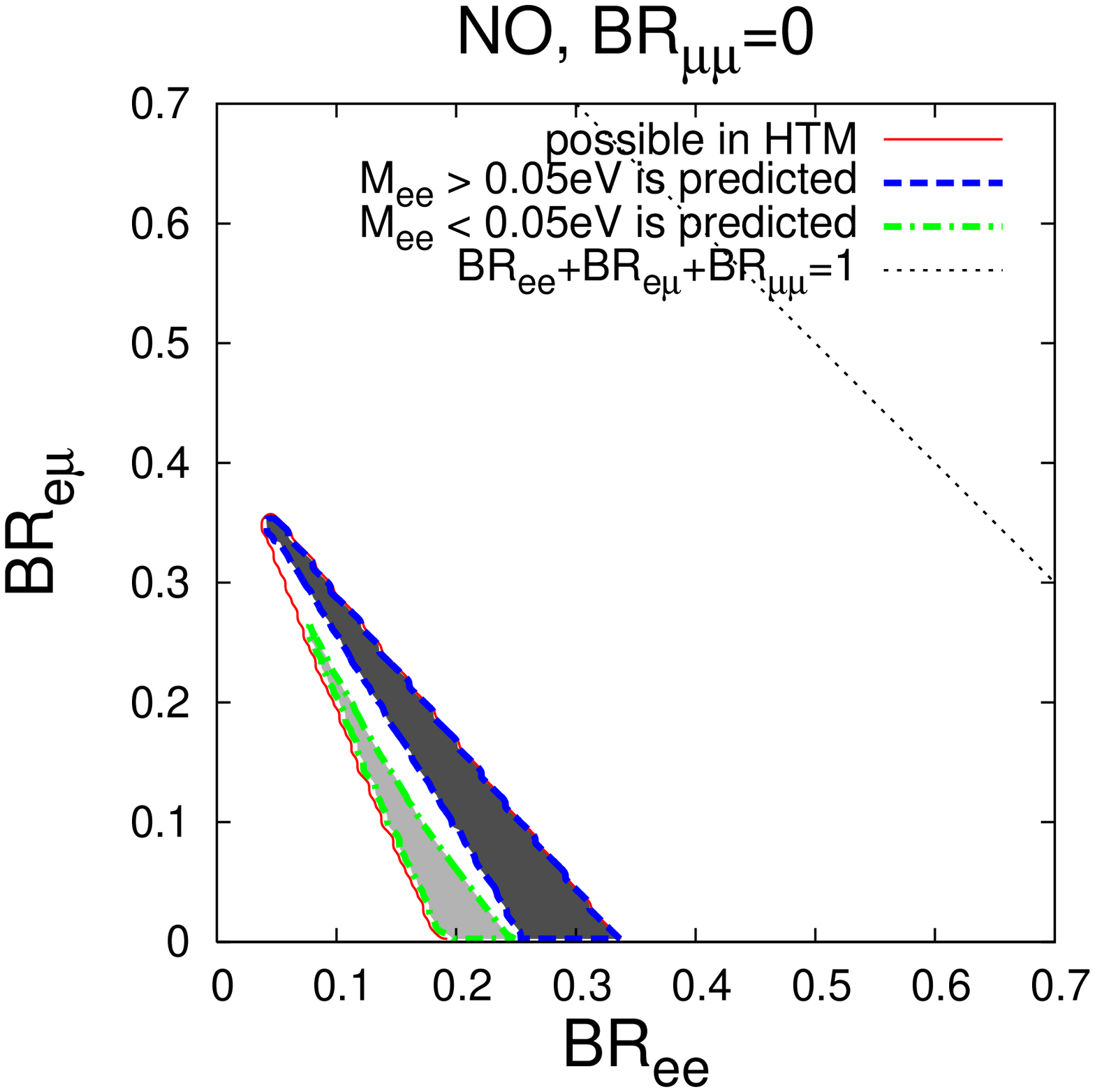}
\includegraphics[angle=-90,width=5cm]{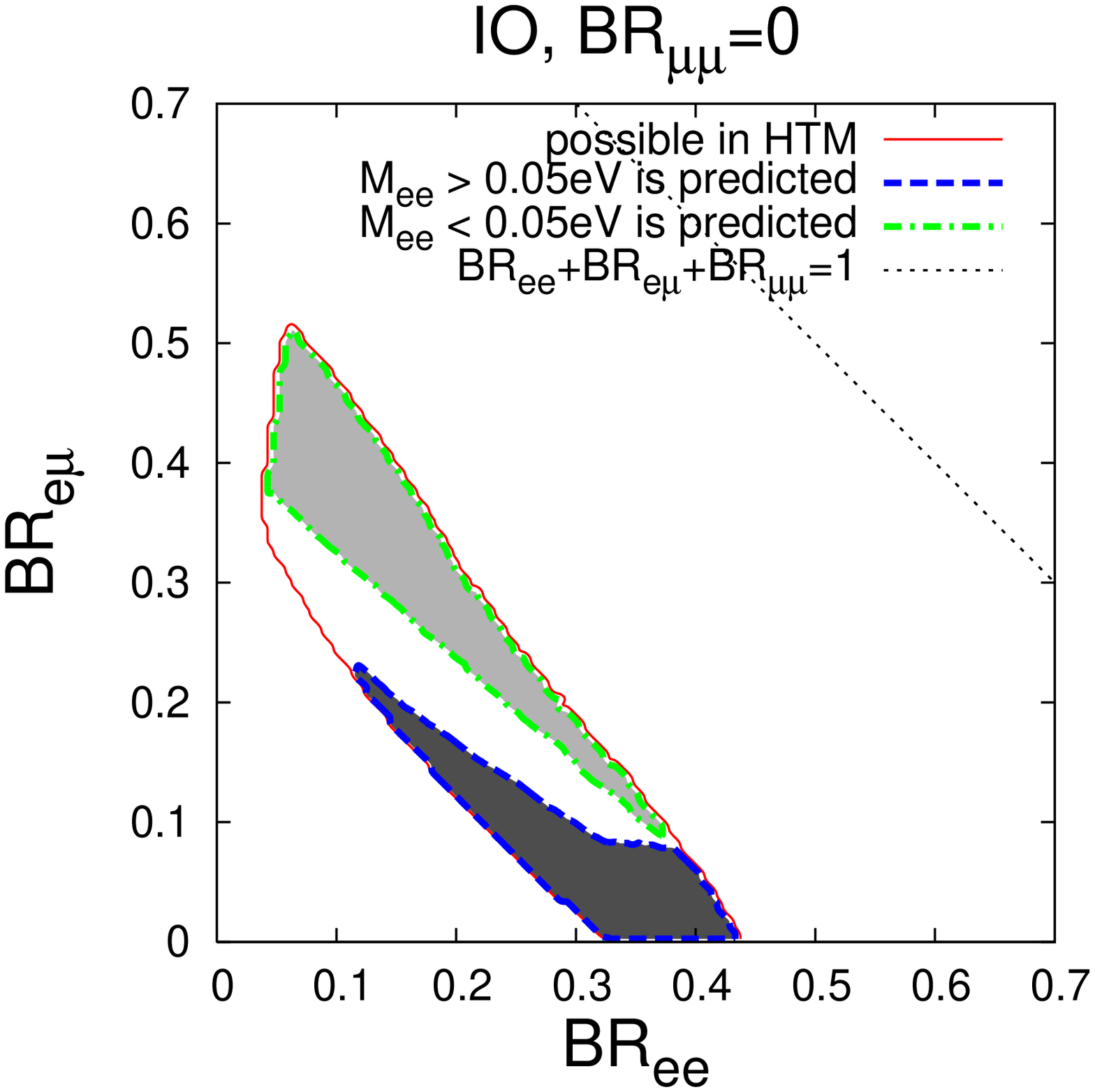}
\includegraphics[angle=-90,width=5cm]{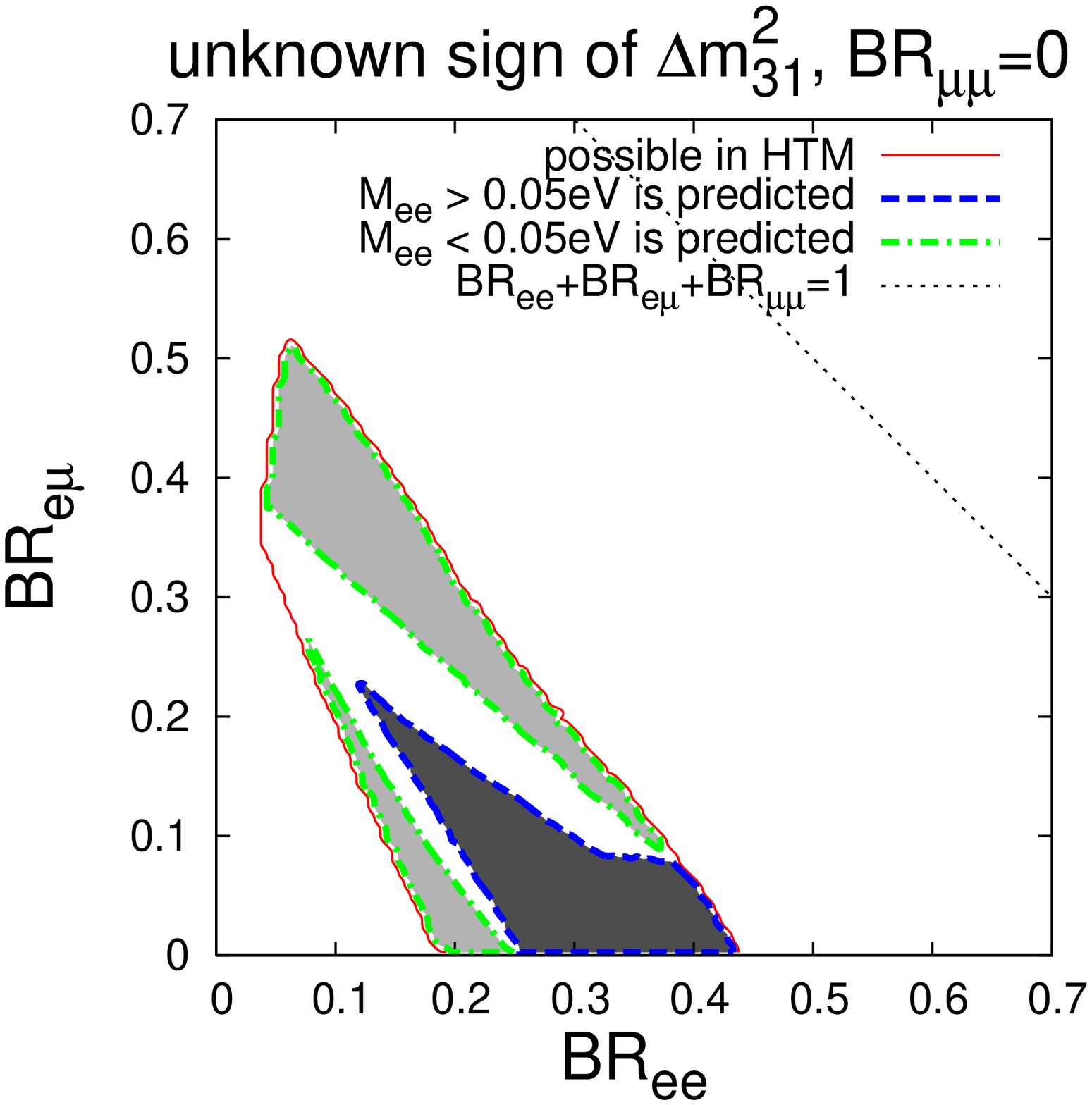}
\caption{The same as in Fig.~\ref{fig:noBRmmimpr}, but assuming that
  the experimentally determined $\BR_{\mu\mu}=0$. The dotted line
  corresponds to $\BR_{ee}+\BR_{e\mu}+\BR_{\mu\mu} = 1$; the region
  above the line is unphysical.}
\label{fig:BRmm0impr}
\end{center}
\end{figure}
%%%%%%%%%%%%%%%%%%%%%%%%%%%%%%%%%%%%%%%%%%%%%
% \begin{figure}
% \begin{center}
% \includegraphics[angle=-90,width=5cm]{BR_eeem_mm01_mee005+-_impr_n.eps}
% \includegraphics[angle=-90,width=5cm]{BR_eeem_mm01_mee005+-_impr_i.eps}
% \includegraphics[angle=-90,width=5cm]{BR_eeem_mm01_mee005+-_impr_ni.eps}
% \caption{
%  The same as in Fig.~\ref{fig:BRmm0impr}, but
% for $\BR_{\mu\mu}=0.1$.
% % is assumed as an extra observation.
% }
% \label{fig:BRmm01impr}
% \end{center}
% \end{figure}
%%%%%%%%%%%%%%%%%%%%%%%%%%%%%%%%%%%%%%%%%%%
\begin{figure}
\begin{center}
\includegraphics[angle=-90,width=5cm]{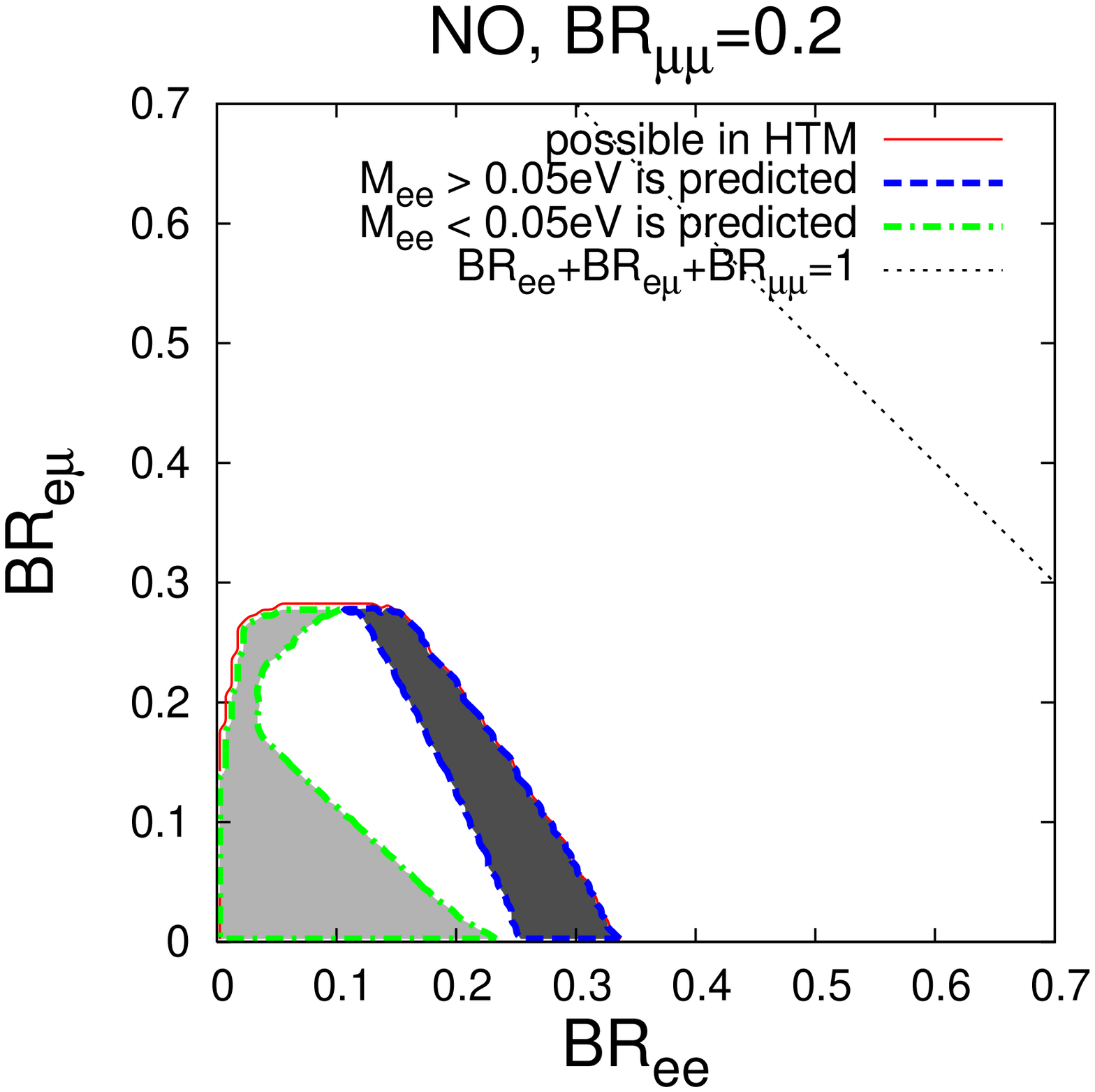}
\includegraphics[angle=-90,width=5cm]{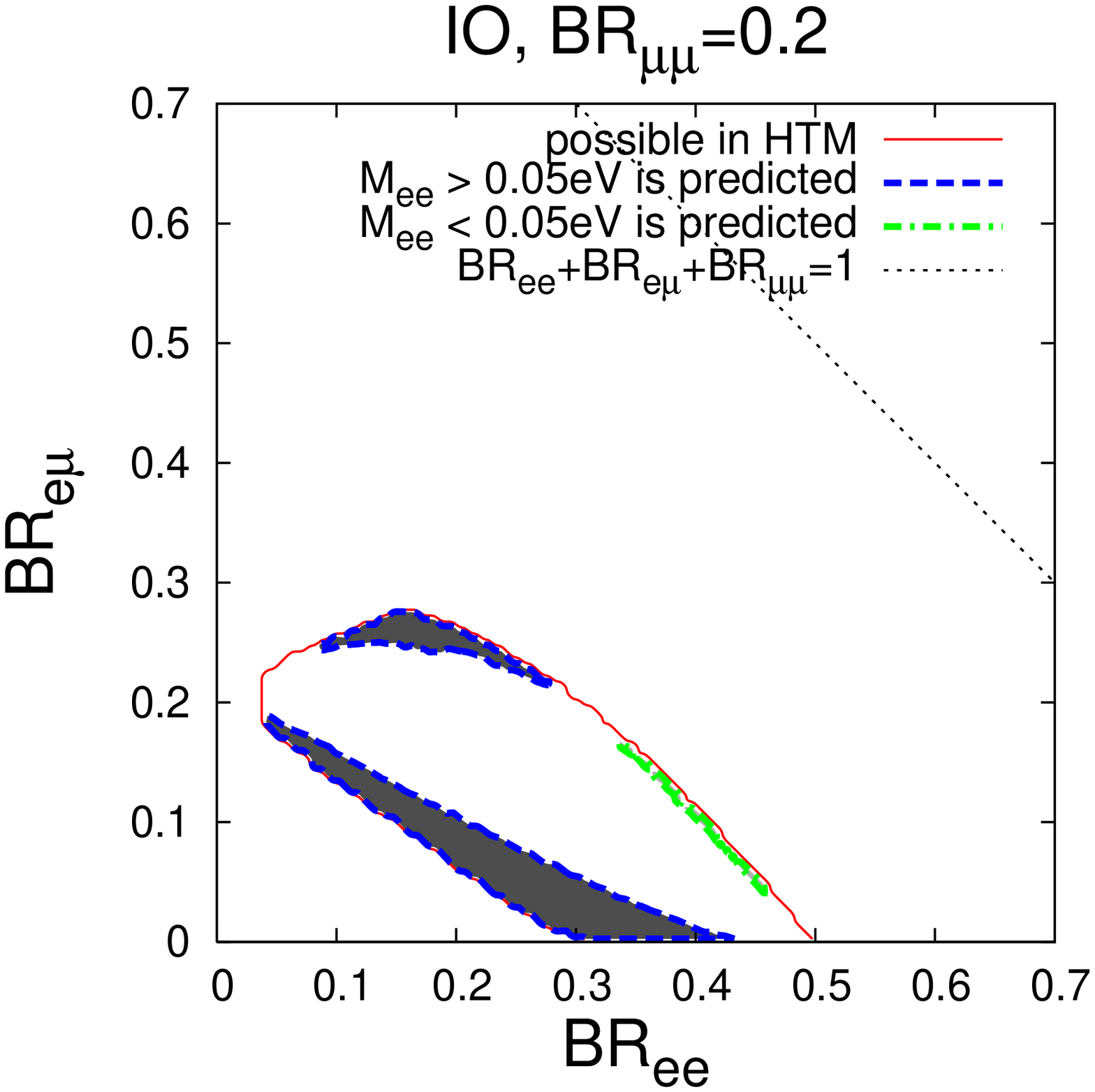}
\includegraphics[angle=-90,width=5cm]{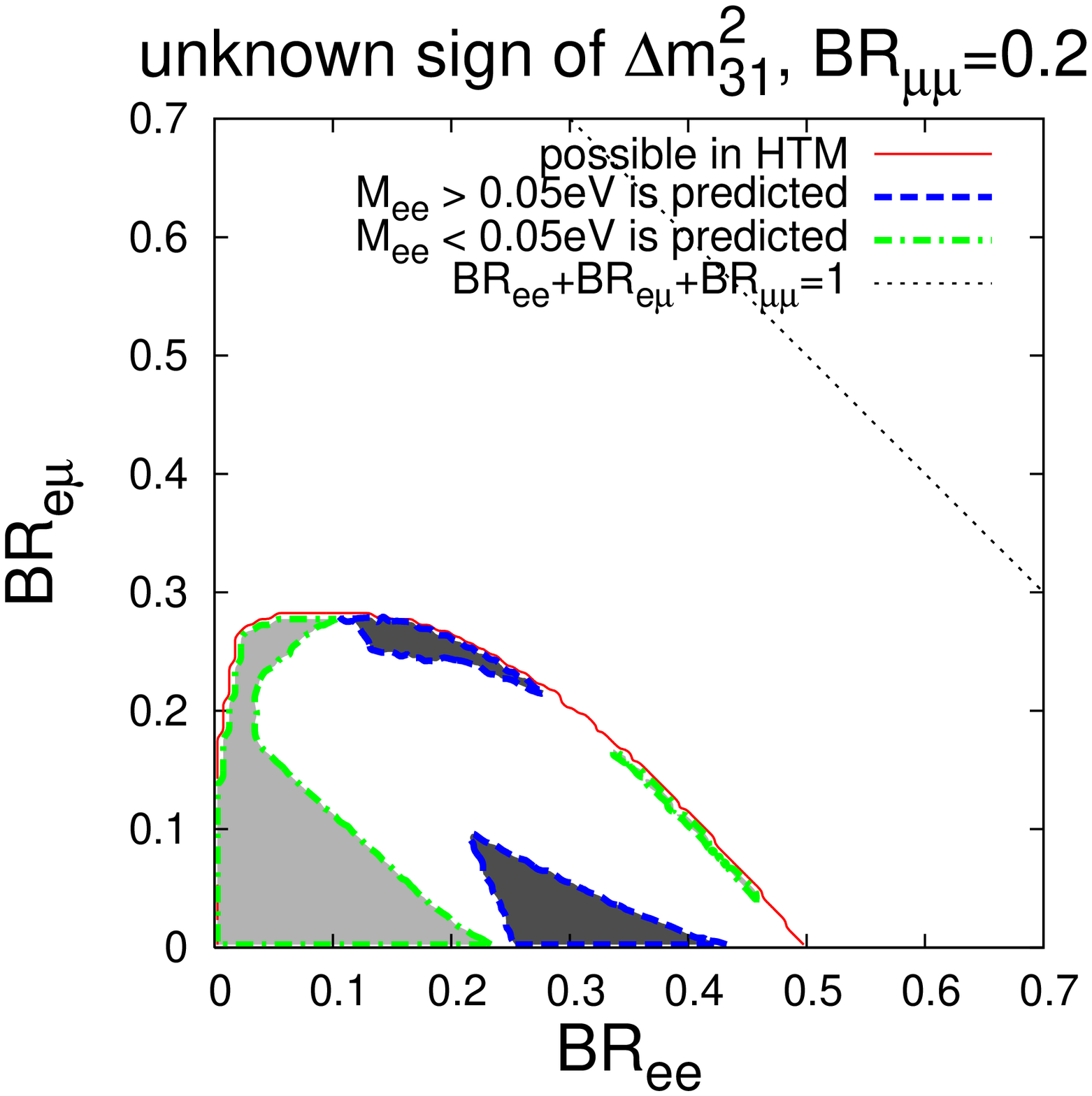}
\caption{The same as in Fig.~\ref{fig:BRmm0impr}, but for $\BR_{\mu\mu}=0.2$.}
\label{fig:BRmm02impr}
\end{center}
\end{figure}
%%%%%%%%%%%%%%%%%%%%%%%%%%%%%%%%%%%%%%%%%%
\begin{figure}
\begin{center}
\includegraphics[angle=-90,width=5cm]{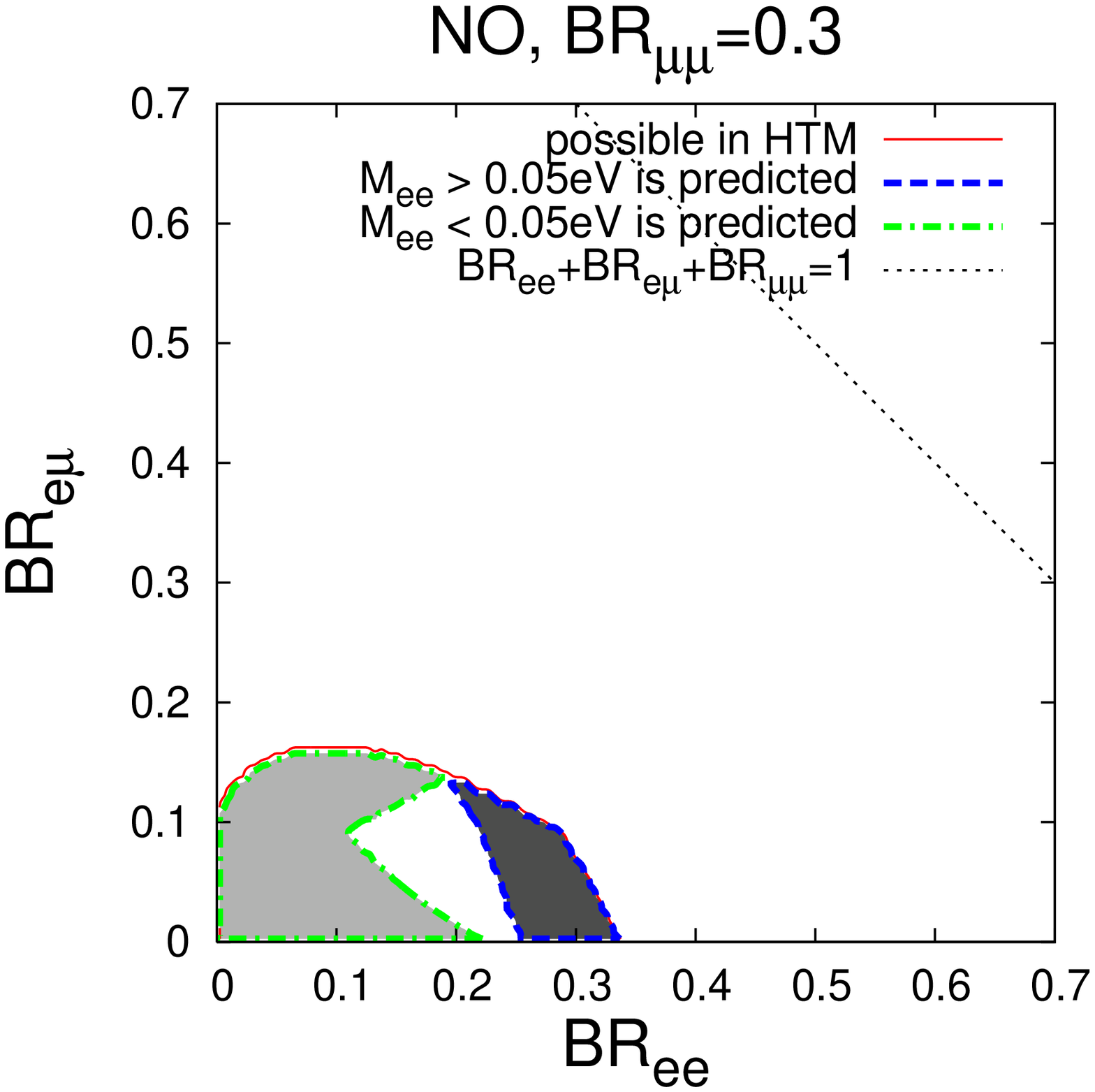}
\includegraphics[angle=-90,width=5cm]{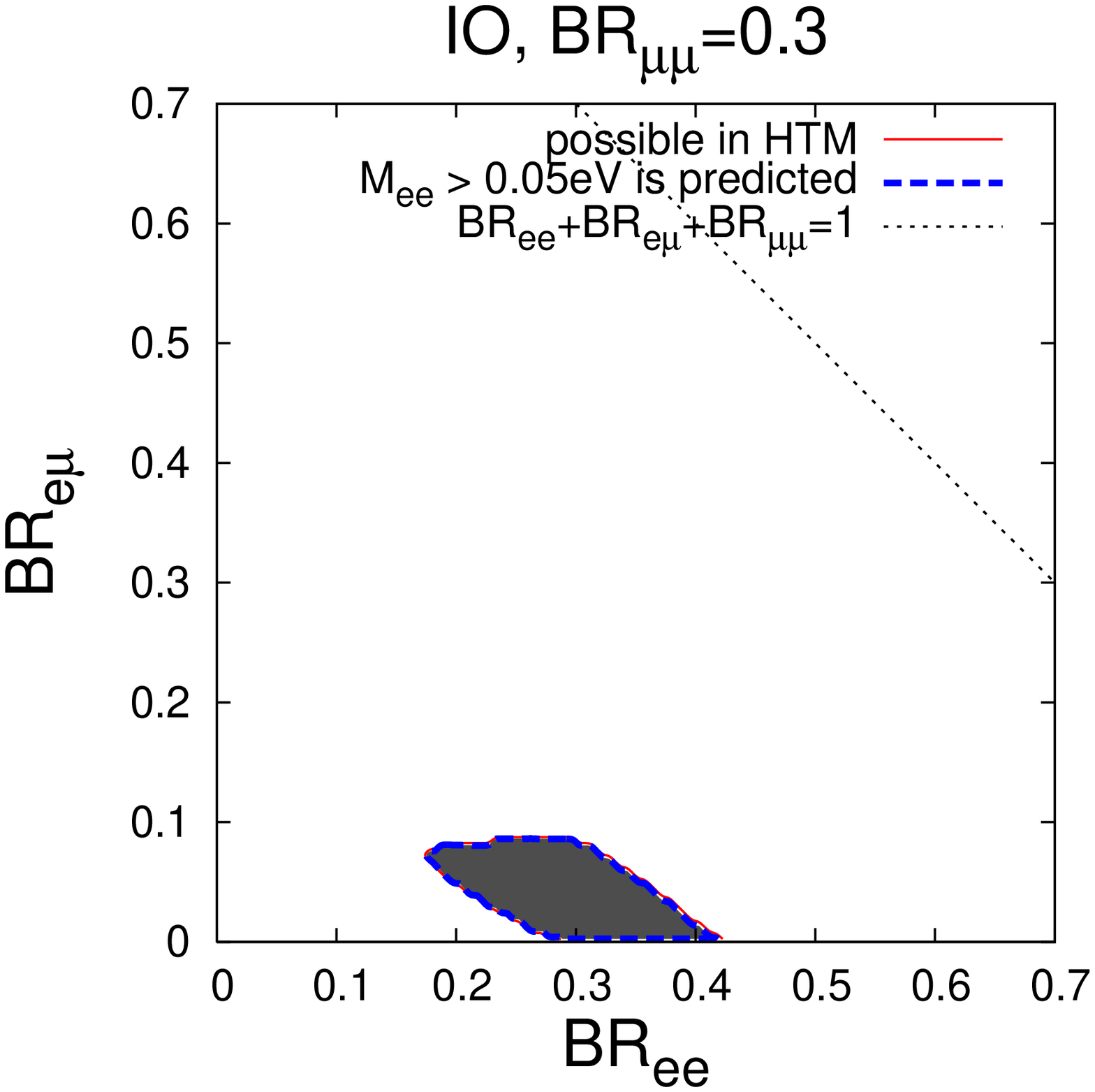}
\includegraphics[angle=-90,width=5cm]{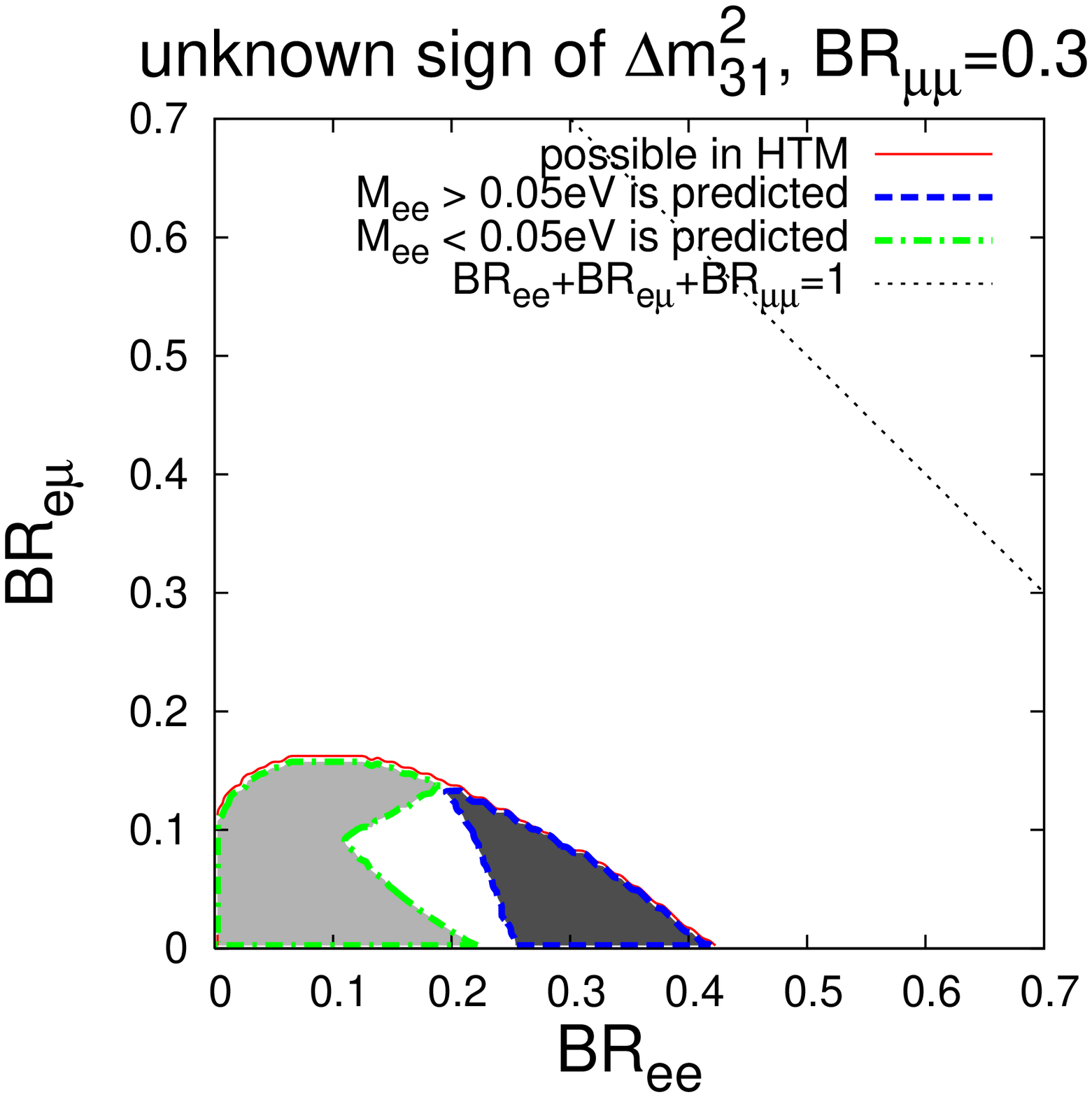}
\caption{The same as in fig.~\ref{fig:BRmm0impr}, but for $\BR_{\mu\mu}=0.3$.}
\label{fig:BRmm03impr}
\end{center}
\end{figure}
%%%%%%%%%%%%%%%%%%%%%%%%%%%%%%%%%%%%
%
\noindent specific value of $\BR_{\mu\mu}=x$, we include an
uncertainty of $\pm 0.01$ in $x$, i.e.\ we use $\BR_{\mu\mu}= x\pm0.01$ 
in the numerical calculations.  The dotted lines in
Figs.~\ref{fig:BRmm0}-\ref{fig:BRmm03} correspond to
$\BR_{ee}+\BR_{e\mu}+\BR_{\mu\mu}=1$.  We do not use this constraint:
the line represents the boundary of the physical region of values of
$\BR_{ee}$, $\BR_{e\mu}$ and $\BR_{\mu\mu}$.

It is clear from Fig.~\ref{fig:BRmm0} that the measurement of
$\BR_{\mu\mu}$ can improve the predictability of $|M_{ee}|$: the
relative magnitude of the white ``degeneracy'' region, in general, is
smaller than in the case when no information on $\BR_{\mu\mu}$ is
available.  This is not the case, however, for IO spectrum and values
of $\BR_{\mu\mu} = 0.1;~ 0.2$ (Figs.~\ref{fig:BRmm01} and \ref{fig:BRmm02}, 
middle panels).

Figures~\ref{fig:noBRmm}-\ref{fig:BRmm03} show also the regions
(dash-dotted line) where one {\it definitely} has $|M_{ee}| <0.05~\eV$.  
If the measured values of $\BR_{ee}$, $\BR_{e\mu}$ and
$\BR_{\mu\mu}$ lie in one of these regions, the observation of
$\betabeta$-decay in the next generation of experiments can be
extremely challenging.  Even in such a case, however, searches for the
$\betabeta$-decay are important and necessary also as a test of the
HTM itself. If the $\betabeta$-decay is observed while the measured
values of the $H^{\pm\pm}$ leptonic decay branching ratios imply,
e.g. a negative result of the searches for $\betabeta$-decay, we will
be led to conclude that $M_{l'l}$ and $h_{l'l}$ are not directly
related: $M_{l'l}\neq \sqrt{2}h_{l'l}v_{\Delta}$.  Such a situation
can arise, for instance, if $v_{\Delta}=0$, or in models with $H^{++}$
which is not an $SU(2)_L$ triplet, but, e.g.\ is a $Y=4$
singlet~\cite{I0Y4} with couplings to the charged leptons given by
$h_{l'l}\, \overline{(l'_{R})^C}\, l_{R}\, H^{++}$.

We have performed the same analysis, but with reduced uncertainties in
$\sin^2{2\theta_{23}}$ and $\sin^2{2\theta_{13}}$:
$\sin^2{2\theta_{23}} > 0.99$ and $\sin^2{2\theta_{13}} < 0.04$.  The
indicated precisions (or better ones) in the determination of
$\sin^2{2\theta_{23}}$ and $\sin^2{2\theta_{13}}$ are expected to be
achieved in the upcoming T2K~\cite{T2K} and reactor antineutrino
experiments Double CHOOZ~\cite{DCHOOZ}, Daya Bay~\cite{DayaB} and
RENO~\cite{RENO}, respectively.  The results of the analysis are shown
graphically in Figs.~\ref{fig:noBRmmimpr}-\ref{fig:BRmm03impr}.  The
notations in Figs.~\ref{fig:noBRmmimpr}-\ref{fig:BRmm03impr} are the
same as in Figs.~\ref{fig:noBRmm}-\ref{fig:BRmm03}. We see from
Figs.~\ref{fig:noBRmmimpr}-\ref{fig:BRmm03impr} that improving the
precision on $\sin^2{2\theta_{23}}$ and $\sin^2{2\theta_{13}}$ leads
to a noticeable reduction of the regions of values of $\BR_{ee}$ and
$\BR_{e\mu}$, for which it is impossible to determine whether
$|M_{ee}| \geq 0.05~\eV$ or $|M_{ee}| < 0.05~\eV$. As
Fig.~\ref{fig:BRmm0impr} demonstrates, the reduction will be
particularly significant if the measured $\BR_{\mu\mu} < 10^{-2}$
(which we remind the reader corresponds to the case denoted by us as
$\BR_{\mu\mu} = 0$).

%%%%%%%%%%%%%%%%%%%%%%%%%%%%%%%%%%%%%%%%%%%%%%%%
%
\section{Conclusions}
%
%%%%%%%%%%%%%%%%%%%%%%%%%%%%%%%%%%%%%%%%%%%%%%%
%

We have investigated the connection between the $\betabeta$-decay
effective Majorana mass $|M_{ee}|$, and the branching ratios of the
decays $H^{\pm\pm}\to l^\pm {l'}^\pm$, $l,{l'} = e,\mu$, of the doubly
charged Higgs boson $H^{\pm\pm}$ within the Higgs Triplet Model (HTM)
of neutrino mass generation.  Our analysis was performed within the
version of the model with explicit breaking of the total lepton charge
conservation, in which $H^{\pm\pm}\to l^\pm {l'}^\pm$, $l,{l'} =
e,\mu,\tau$, are the dominant decay modes of $H^{\pm\pm}$.  In this
model the couplings of the doubly charged Higgs field $H^{++}$ to the
flavour neutrinos and charged leptons are proportional to the elements
of the Majorana mass matrix of the (flavour) neutrinos, $M_{{l'}l}$,
and the branching ratios $\BR(H^{\pm\pm}\to l^\pm {l'}^\pm)$ are
entirely determined by the elements of the PMNS matrix and neutrino
masses. The latter possibility is realised if the mass of the doubly
charged Higgs scalar does not exceed the mass of the singly charged
one, $M_{H^{\pm\pm}} \leq M_{H^\pm}$, and if the vacuum expectation
value of the neutral component of the Higgs triplet field satisfies
$v_\Delta \lesssim 1~\text{MeV}$.  The model under discussion was
shown~\cite{Akeroyd:2007zv,Garayoa:2007fw,Kadastik:2007yd} to have a
rich and physically interesting phenomenology owing to the fact that
the physical doubly charged and singly charged Higgs fields,
$H^{\pm\pm}$ and $H^{\pm}$, can have masses in the range from
$\sim$100~GeV to $\sim$1~TeV and thus can, in principle, be produced
and observed at LHC\@.  More importantly, by studying the decays
$H^{\pm\pm}\to l^\pm {l'}^\pm$, $l,{l'} = e,\mu,\tau$, it might be
possible to obtain information on the absolute neutrino mass scale, on
the type of neutrino mass spectrum and on the Majorana CP violating
phases present in the neutrino mixing matrix~\cite{Akeroyd:2007zv,Garayoa:2007fw,Kadastik:2007yd}.

In the present article we have investigated the possibility to use the
information on the neutrino mass spectrum and the Majorana CP
violating phases from the measurements of the $H^{\pm\pm}\to l^\pm{l'}^\pm$ 
decay branching ratios, $\BR(H^{\pm\pm}\to l^\pm {l'}^\pm)$,
$l,{l'} = e,\mu$, in order to obtain predictions for the effective
Majorana mass in neutrinoless double beta ($\betabeta$-) decay,
$|M_{ee}|$.  Among the different decay channels $H^{\pm\pm}\to l^\pm{l'}^\pm$, 
$l,{l'} = e,\mu,\tau$, the easier to observe and measure
the corresponding branching ratios with high precision are those with
two electrons (positrons), two muons (antimuons), or an electron
(positron) and a muon (antimuon), $e^{\pm}e^{\pm}$, $e^{\pm}\mu^{\pm}$
and $\mu^{\pm}\mu^{\pm}$, in the final state. If the mass of
$H^{\pm\pm}$ does not exceed approximately 400~GeV, the branching
ratios of the $H^{\pm\pm}$ decays into $e^{\pm}e^{\pm}$,
$e^{\pm}\mu^{\pm}$ and $\mu^{\pm}\mu^{\pm}$ can be measured at LHC
with a few percent error~\cite{Perez:2008ha}.

Taking into account the current and prospective uncertainties in the
values of the neutrino mixing parameters most relevant for the problem
studied - the atmospheric neutrino mixing angle $\theta_{23}$ and the
CHOOZ angle $\theta_{13}$, and allowing the lightest neutrino mass and
the CP violating Dirac and Majorana phases to vary in the intervals
$[0, 0.3~\eV]$ and $[0, 2\pi]$, respectively, we have derived the regions
of values of $\BR(H^{\pm\pm}\to e^\pm e^\pm)$ and $\BR(H^{\pm\pm}\to e^\pm \mu^\pm)$ 
for which we {\it definitely} have $|M_{ee}|\geq0.05$~eV, or $|M_{ee}|<0.05$~eV. This 
is done for neutrino mass
spectrum with normal ordering (NO), inverted ordering (IO) and in the
case when the type of the spectrum is not known. In what concerns the
branching ratio $\BR(H^{\pm\pm}\to \mu^\pm \mu^\pm)$, we have
considered two cases: i) the possible data on $\BR(H^{\pm\pm}\to\mu^\pm \mu^\pm)$ 
is not used as an additional constraint in the
analysis, ii) the possible data on $\BR(H^{\pm\pm}\to \mu^\pm\mu^\pm)$ is 
included in the analysis.  In the latter case, results
for several values of $\BR(H^{\pm\pm}\to \mu^\pm \mu^\pm)$ have been
obtained.

Our results are presented graphically in
Figs.~\ref{fig:noBRmm}-\ref{fig:BRmm03impr}. They show that if the
doubly charged Higgs bosons $H^{\pm\pm}$ will be discovered at LHC and
at least the two branching ratios $\BR(H^{\pm\pm}\to e^\pm e^\pm)$ and
$\BR(H^{\pm\pm}\to e^\pm \mu^\pm)$ will be measured with a sufficient
accuracy, one can obtain important information on the
$\betabeta$-decay effective Majorana mass $|M_{ee}|$.  In the various
cases considered, we have identified the regions values of
$\BR(H^{\pm\pm}\to e^\pm e^\pm)$ and $\BR(H^{\pm\pm}\to e^\pm\mu^\pm)$, 
for which $|M_{ee}|$ is {\it definitely} bigger or smaller
than 0.05~eV (Fig.~\ref{fig:noBRmm}).  We have shown also that due to
i) the uncertainties in the determination of $\sin^22\theta_{23}$ and
$\sin^22\theta_{13}$, ii) the absence of data on the CP violating
phases in the neutrino mixing matrix, and iii) the existing rather
loose upper bound on the absolute neutrino mass scale, there exist
also noticeable regions of values of $\BR(H^{\pm\pm}\to e^\pm e^\pm)$
and $\BR(H^{\pm\pm}\to e^\pm \mu^\pm)$ for which it is impossible to
determine unambiguously whether $|M_{ee}| \geq 0.05~\eV$ or $|M_{ee}|< 0.05~\eV$ 
(Fig.~\ref{fig:noBRmm}).  This ``degeneracy'' can be
partially lifted by using the additional information from a
measurement of $\BR_{\mu\mu}$ (Figs.~\ref{fig:BRmm0}-\ref{fig:BRmm03}).

The same analysis was performed with reduced uncertainties in
$\sin^2{2\theta_{23}}$ and $\sin^2{2\theta_{13}}$ corresponding to
$\sin^2{2\theta_{23}} > 0.99$ and $\sin^2{2\theta_{13}} < 0.04$.  The
results are presented graphically in
Figs.~\ref{fig:noBRmmimpr}-\ref{fig:BRmm03impr}.  They show that
improving the precision on $\sin^2{2\theta_{23}}$ and
$\sin^2{2\theta_{13}}$ leads to a noticeable reduction of the regions
of values of $\BR_{ee}$ and $\BR_{e\mu}$ for which it is impossible to
determine whether $|M_{ee}| \geq 0.05~\eV$ or $|M_{ee}| < 0.05~\eV$.
The reduction will be particularly significant if the measured
$\BR_{\mu\mu} < 10^{-2}$. \\%[5mm]

\section*{Acknowledgements.}

This work was supported in part by the INFN under the program ``Fisica
Astroparticellare'', by the European Network of Theoretical
Astroparticle Physics ILIAS/N6 (contract RII3-CT-2004-506222) and by
World Premier International Research Center Initiative (WPI
Initiative), MEXT, Japan.  S.T.P. acknowledges with gratefulness the
hospitality and support of IPMU, University of Tokyo, where part of
the work on the present article was done.

\end{document}